\documentclass[%
preprint,
 amsmath,amssymb,
 aps,
]{revtex4-1}

\usepackage{graphicx}
\usepackage{dcolumn}
\usepackage{bm}

\usepackage{lineno}
\usepackage{longtable}

\begin{document}

\title[Extremes and reversibility]{Extremes, intermittency and time reversibility of atmospheric  turbulence at the cross-over from production to inertial scales}%

\author{E. Zorzetto}
 \email{enrico.zorzetto@duke.edu.}
 \affiliation{Division of Earth and Ocean Sciences, Nicholas School of the Environment, Duke University, Durham, North Carolina 27708, USA}
\author{A.D. Bragg}%
\email{andrew.bragg@duke.edu.}
\affiliation{ 
Department of Civil and Environmental Engineering, Duke University, Durham, North Carolina 27708, USA 
}%

\author{G. Katul}
 \email{gaby@duke.edu.}
\affiliation{Nicholas School of the Environment, Duke University, Durham, North Carolina 27708, USA and The Karlsruher Institut f{\rm$\ddot{u}$}r Technologie (KIT)/IMK-IFU, Kreuzeckbahnstra{\rm$\beta$}e 19, 82467 Garmisch-Partenkirchen, Germany 
}%

\date{\today}

\begin{abstract}

The effects of mechanical generation of turbulent kinetic energy and buoyancy forces on the statistics of air temperature and velocity increments are experimentally investigated at the cross over from production to inertial range scales. The ratio of an approximated mechanical to buoyant production (or destruction) of turbulent kinetic energy can be used to form a dimensionless stability parameter $\zeta$ that classifies the state of the atmosphere as common in many atmospheric surface layer studies. Here, we assess how $\zeta$ affects the scale-wise evolution of the probability of extreme air temperature excursions, their asymmetry and time reversibility. The analysis makes use of high frequency velocity and air temperature time series measurements collected at $z$=5 m above a grass surface at very large frictional Reynolds numbers $Re_*=u_* z/\nu > 1\times 10^5$ ($u_*$ is the friction velocity and $\nu$ is the kinematic viscosity of air). Using conventional higher-order structure functions, temperature exhibits larger intermittency and wider multifractality when compared to the longitudinal velocity component, consistent with laboratory studies and simulations conducted at lower $Re_*$. Moreover, deviations from the classical Kolmogorov scaling for the longitudinal velocity are shown to be reasonably described by the She-Leveque vortex filament model that has no 'tunable' parameters and is independent of $\zeta$. The work demonstrates that external boundary conditions, and in particular the magnitude and sign of the sensible heat flux, have a significant impact on temperature advection-diffusion dynamics within the inertial range. In particular, atmospheric stability affects both the buildup of intermittency and the persistent asymmetry and time irreversibility observed in the first two decades of inertial sub-range scales.

\end{abstract}

\keywords{Atmospheric Boundary Layer,  Extreme Value Statistics, Turbulence,  Time-reversibility, Ramp-Cliff patterns}
\maketitle

\section{Introduction}
Turbulence in fluids is prototypical of spatially extended nonlinear dissipative systems characterized by large fluctuations that are active over wide ranging scales \cite{sreenivasan1999fluid}.  Scalar turbulence is by no means an exception to this description. Scalar turbulence share many phenomenological parallels with the much studied turbulent velocity fluctuations, especially in the inertial subrange. However, scalar turbulence also exhibits distinctive large- and fine-scaled temporal patterns (e.g. ramp-cliff) that are usually weak or all together absent from their component-wise turbulent velocity counterparts \cite{antonia1979temperature,shraiman2000scalar,warhaft2000passive}. This finding is particularly true in the atmospheric surface layer (ASL) \cite{garratt1994review,stull2012introduction}, a layer within the atmospheric boundary layer (ABL) that is sufficiently far above roughness elements but not too far from the ground to be directly impacted by the Coriolis force.  In the ASL, the frictional Reynolds number $Re_* = u_* z/\nu$ can readily exceed $10^5$, where $z$ is the distance above the ground surface, $u_*$ is the friction velocity related to the kinematic turbulent stress, and $\nu$ is the kinematic viscosity of air.  A direct consequence of this large $Re_*$ is a wide separation between scales over which turbulent kinetic energy ($k$) is produced and dissipated. In the absence of thermal stratification, $k$ is produced at scales commensurate with $z$; however, the action of fluid viscosity responsible for the dissipation of $k$ occurs at scales commensurate to or smaller than the Kolmogorov microscale $\eta_K=(\nu^3/ \langle \epsilon \rangle)^{1/4}$, where $\langle \epsilon \rangle$ is the mean turbulent kinetic energy dissipation rate that is proportional to $u_*^3/z$ for a neutrally stratified ASL \cite{stull2012introduction}. These estimates of $\langle \epsilon \rangle$ and $\eta_K$ result in $z/\eta_K \sim Re_*^{3/4} > 5000$ in the ASL, which is rarely achieved in direct numerical simulations or laboratory studies. Embedded in this wide ranging scale separation is the inertial subrange \cite{kolmogorov1941local}, where self similar scaling of velocity and air temperature structure functions is expected to hold for eddy sizes much larger than $\eta_K$ but much smaller than $z$. Integral scales or scales comparable to $z$ are directly influenced by boundary conditions imposed on the flow including surface heating (or cooling) in the ASL, whereas small scales (e.g. $\eta_K$) may attain universality and local isotropy after a large number of cascading steps away from the energy injection scales. 

Much attention has been historically dedicated to the inertial subrange and the subsequent cross-over to the viscous or molecular regimes precisely because of the possible universal character of turbulence at such fine scales \cite{kraichnan1968small,kuznetsov1992fine,warhaft2000passive,schumacher2014small,yeung2015extreme,katul2015bottlenecks}.  However, it is now accepted that some coupling between small and large scales exists, especially for passive scalars \cite{warhaft2000passive,shraiman2000scalar,katul2006analysis}, that act to enhance intermittency buildup across scales and distort any universal behavior by injecting the effects of the boundary conditions (or the $k$ generation mechanism). Along similar lines of inquiry, it has been conjectured that the presence of coherent ramp-cliff patterns in concentration (or temperature) time series are responsible, to some degree, for this coupling \cite{warhaft2000passive}. Ramp-cliff structures are characterized by local intense scalar gradients separated by large quiescent regions. The presence of ramp-cliff structures in scalar time series has been shown to break locality of eddy interactions and determine some departures from small scale isotropy. 

Sweep-ejection dynamics connected to the presence of ramps are likely to play a major role in observed extreme value statistics, as shown e.g., for Lagrangian velocity sequences in plant canopy turbulence \citep{reynolds2012gusts}. Moreover, ramps are asymmetric and produce non-zero odd ordered structure functions, sharing striking resemblance with flight-crash events recently reported for the turbulent kinetic energy of Lagrangian particles \cite{xu2014flight}. Even though ramps have been extensively observed experimentally \citep{antonia1979temperature}, studied as surface renewal processes \cite{katul2006analysis}, and from a Lagrangian perspective \cite{shraiman2000scalar,falkovich2001particles}, a unified picture on their effects on inertial scales statistics remains lacking and motivates the work here.

Our main objective is to investigate two questions about scalar turbulence at scales spanning production to inertial subranges: How do ramp-cliff patterns modify (i) the probability of extreme scalar concentration excursions and its corollary intermittency buildup, and (ii) symmetry and time reversibility of scalar turbulence. These two questions are explored for differing turbulent energy injection mechanisms (mechanical and buoyancy forces) in the ASL. Here we focus on the production-to-inertial scales instead of the usual inertial to viscous ranges for the following reasons. First, any cross-scale coupling with ramp-cliff patterns is likely to be sensed at large scales commensurate with the ramp durations. Second, these scales are deemed most relevant when constructing sub-grid scale models for improving Large Eddy Simulations \cite{meneveau1996lagrangian,porte2000scale,higgins2003alignment,stoll2006dynamic}. Third, these scales encode much of the scalar variance that is needed when deriving phenomenological theories for the bulk flow properties based on the spectral shapes of the turbulent velocity and air temperature \cite{katul2011mean,katul2013mean,li2012mean,katul2014two,li2015revisiting}, especially for the ASL. 

To achieve the study objectives, high frequency measurements of the three velocity components and air temperature fluctuations in the ASL are used to explore flow statistics at the transition from production to inertial scales. In particular, the focus is on the first two decades dominated by approximate inertial subrange effects, where the transition from the large eddies to the universal equilibrium or inertial range occurs. The statistical properties of temperature increments within this range of scales is examined with the goal of addressing to what extent the tail properties (and thus the probability of extreme events) at fine scales still carry signatures from the production ranges and in particular of large coherent structures such as ramp-cliffs.  The experiments here spanned several atmospheric stability regimes that dictate to what degree turbulent kinetic energy is mechanically or buoyantly generated (or dissipated) depending on surface heating (or cooling) and on the turbulent shear stress near the ground \cite{monin1954basic}. However, due to the large Reynolds number in our experimental setting, the stratification is not sufficiently severe to allow for a transition to non-turbulent regimes. Therefore, the turbulence can be studied as three dimensional and fully developed.

\section{Theory}
\subsection{Overview of ASL similarity at large- and small-scales}
The turbulent kinetic energy budget for a stationary and planar homogeneous flow in the absence of subsidence is given by
\begin{equation}
\label{TKE_Budget}
\frac{\partial k}{\partial t_0}=0=-\overline{u'w'} \frac{dU}{dz}+{\beta_o g} \overline{w'T'}+P_D+T_T-\epsilon,
\end{equation}
where $k=(\overline{u'^2 + v'^2 + w'^2})/2$ is the turbulent kinetic energy, $u'$, $v'$, and $w'$ are the turbulent velocity components along the mean wind (or $x$), lateral (or $y$), and vertical (or $z$) directions, respectively, $t_0$ is time, and the five terms on the right-hand side of Eq. (\ref{TKE_Budget}) are mechanical production, buoyant production (or destruction), pressure transport, turbulent transport of $k$, and viscous dissipation of $k$, respectively, $\beta_o$ is the thermal expansion coefficient for gases ($\beta_o=1/T$, $T$ is air temperature here), $g$ is the gravitational acceleration, $-\overline{u'w'}=u_*^2$ is the turbulent kinematic shear stress near the surface, and $\overline{w'T'}$ is the kinematic sensible heat flux from (or to) the surface.  When $\overline{w'T'}>0$, buoyancy is responsible for the generation of $k$ and the ASL is classified as unstable.  When $\overline{w'T'}<0$, the ASL is classified as stable and buoyancy acts to diminish the mechanical production of $k$.  The relative significance of the mechanical production to the buoyancy generation (or destruction) may be expressed as
\begin{equation}
\label{TKE_Production}
-\overline{u'w'} \frac{dU}{dz}+{\beta_o g} \overline{w'T'}=\frac{u_*^3}{\kappa z} \left[\phi_m(\zeta) +\frac{\kappa z {\beta_o g} \overline{w'T'}}{u_*^3}\right]=\frac{u_*^3}{\kappa z} \left[\phi_m(\zeta) -\zeta\right],
\end{equation}
where
\begin{equation}
\label{TKE_Production2}
\frac{dU}{dz}= \frac{u_*}{\kappa z} \phi_m(\zeta), \quad \zeta=\frac{z}{L}, \quad L=-\frac{u_{*}^3}{\kappa g \beta_o \overline{w'T'}},
\end{equation}
and $\phi_m(\zeta)$ is known as a stability correction function reflecting the effects of thermal stratification on the mean velocity gradient ($\phi_m(0)=1$ recovers the von Karman-Prandtl log-law), $\kappa \approx 0.4$ is the von Karman constant, and $L$ is known as the Obukhov length as described by the Monin and Obukhov similarity theory \cite{monin1954basic}.  The physical interpretation of $L$ is that it is the height at which mechanical production balances the buoyant production or destruction when $\phi_m(\zeta)$ does not deviate appreciably from unity.  For a neutrally stratified ASL flow, $|L| \rightarrow \infty$ and $|\zeta| \rightarrow 0$. The sign of $L$ reflects the direction of the heat flux, with negative values of $L$ corresponding to upward heat fluxes (unstable atmospheric conditions) and positive values $L$ corresponding to downward heat flux (stable atmosphere).

Several bulk flow statistics in the ASL can be reasonably described by the aforementioned Monin-Obukhov similarity theory, including the mean air temperature gradient $dT/dz$ and the air temperature variance $\overline{T'^2}$, both varying with $\zeta$ when normalized by a temperature scale $T_*=-\overline{w'T'}/u_*$.  However, the statistics of large-scale features within the temperature time series traces such as the statistics of ramp-cliff patterns do not scale with $z$.  For starters, the ramp characteristic dimension is generally larger than $z$ and their duration exceeds $(\kappa z \phi_m(\zeta)^{-1}) u_*^{-1}$.  Ramps have been observed within canopies, near the canopy atmosphere interface, and other types of flows as reviewed elsewhere \cite{katul2006analysis,warhaft2000passive}.  However, $z/L$ does indirectly impact several features of the ramp-pattern in air temperature traces sampled within the ASL.  For example, in stably stratified ASL flows, the temperature ramps appear 'inverted' when compared to their near-neutral counterparts. The amplitudes and durations of ramps can increase with increasing instability due to weaker shearing and intense buoyant updrafts \cite{chen1997coherent,thomas2007organised}. 

At small scales associated with the inertial subrange, the velocity and temperature second-order structure functions are commonly described by the Kolmogorov 1941 (K41) theory \cite{kolmogorov1941local} given as
\begin{equation}
\label{SF_MODELS_K41u}
S^2_u (r)=\overline{[\Delta u(r)]^2}=4 C_{o,u} ( \langle \epsilon \rangle r)^{2/3},
\end{equation}
\begin{equation}
\label{SF_MODELS_K41w}
S^2_w (r)=\overline{[\Delta w(r)]^2}=4 C_{o,w} (\langle \epsilon \rangle r)^{2/3},
\end{equation}
\begin{equation}
\label{SF_MODELS_K41T}
S^2_T (r)=\overline{[\Delta T(r)]^2}=4 C_{o,T} \langle \epsilon_{T} \rangle  \langle \epsilon \rangle^{-1/3} r^{2/3},
\end{equation}
where $\Delta u(r)=u(x+r)-u(x)$, $\Delta w(r)=w(x+r)-w(x)$, and $\Delta T(r)=T(x+r)-T(x)$ are the velocity and temperature increments at separation distance (or scale) $r$, $\langle \epsilon \rangle$ and $\langle \epsilon_T \rangle$ are the $k$ and temperature variance dissipation rates respectively, $C_{o,u}$ and $C_{o,w}$ are the Kolmogorov constants for the longitudinal and vertical velocity components, and $C_{o,T}$ is the Kolmogorov-Obukhov-Corrsin (KOC) constant.  These scaling laws, obtained under the assumptions of similarity and local isotropy, appear to hold reasonably in the ASL for scales smaller than $z/2$ \cite{katul1997energy}. Moreover, the normalized third order structure functions
\begin{equation}
\label{Sr}
S(r)=\frac{ S^3_u  }{ \left( S^2_u \right)^{3/2} }=\frac{ \langle \Delta u (r)^3 \rangle }{  \langle \Delta u (r)^2 \rangle^{3/2}}
\end{equation}
and
\begin{equation}
\label{Fr}
F(r)=\frac{ S^3_{TTu}  }{  S^2_T \left[S^2_u \right]^{1/2} }=\frac{ \langle \Delta u (r) \Delta T(r)^2 \rangle }{ \langle  \Delta T(r)^2 \rangle \langle \Delta u (r)^2 \rangle^{1/2}}
\end{equation}
must be constant to recover K41 predictions for $S^2_u$ and $S^2_T$ in the inertial range \cite{obukhov1949local}.

However, relevant deviations from K41 scaling have been reported for higher order structure functions, especially for the scalar fluctuations. These deviations arise as (i) Eqs. (\ref{SF_MODELS_K41u}) - (\ref{SF_MODELS_K41T}) do not account for intermittency related to spatial variability of the actual $\epsilon$ and $\epsilon_{T}$, and (ii) the hypothesis of local isotropy might not hold for scalars due to non-local interactions across scales \cite{sreenivasan1991local}.  A signature of the latter is the large structure skewness for temperature determined by ramp structures \cite{katul1997energy,warhaft2000passive}. Many models, starting from Kolmogorov's log-normal dissipation rate refinement \cite{kolmogorov1962refinement}, seek to relax some of the restrictive assumptions of K41 so as to explain the anomalous scaling observed in higher order moments.  For scalars, these corrections are commonly expressed as

\begin{equation}
\label{scaling_anomaly}
S^n_T= C_n \left( \epsilon r \right)^{n/3} \left( r/L_I \right)^{\zeta'_n-n/3}
\end{equation}
where the exponent $\zeta'_{n}$ implies a scaling different from K41 that depends on the moment order $n$. The presence of an integral time scale $L_I$ suggests an explicit dependence on large scale eddy motion within the inertial subrange. One estimate of $L_I$ may be derived from the integral length scale of the flow given by
\begin{equation}
L_I= U \cdot I_w=U \cdot \int_0^{\infty} \rho_w (\tau_0) d \tau_0,
\label{Iw}
\end{equation}
where $\rho_w (\tau_0)$ is the vertical velocity autocorrelation function and $\tau_0$ is the time lag. Here, $I_w$ is presumed to be the most restrictive scale given that $w'$ is the flow variable most impacted by the presence of the boundary.

The statistics of air temperature increments across scales ($\tau_0/I_w$) for different $\zeta$ conditions are explored with a lens on two primary features: buildup of heavy tails and destruction of asymmetry originating from ramp-cliff structures at the cross-over from $\tau_0/I_w>1$ to $\tau_0/I_w \approx 0.1 $. Because changes in $\zeta$ do result in changes in $I_w$, the time (or space) lags are presented in dimensionless form as $\tau=\tau_0/I_w$, so that the increments of a flow variable $\Delta s$, with $\Delta s=\Delta u, \Delta w,\Delta T$ at a given dimensionless scale $\tau$, can be expressed as $ \Delta s (\tau) = s(t + \tau) - s(t)$, where $t = t_0/I_w$.

\subsection{Probabilistic description of intermittency}

A number of models have been proposed to capture the effects of intermittency on the flow statistics in the inertial range of scales (e.g., lognormal, bi- and multi-fractals - beta model, log-stable, She-Leveque vortex filaments, etc) and documented by several ASL experiments \cite{katul1994intermittency,katul2001estimating}. Common to all these models is the hypothesis of local isotropy and the accounting for uneven distribution of eddy activity in the space/time domain, which explains the anomalous scaling of higher order even structure functions.

Here, a statistical description of scalar increments is used to fingerprint large-scale signatures across scales $\tau$ for different $\zeta$. If such fingerprints exist, the dissipation rates $\epsilon$ and $\epsilon_T$ need not be sufficient to describe all aspects of the inertial range statistics. The one-time probability density function (pdf) of the increments $\Delta s (\tau)$ of the flow variable $ s = u, w, T$ at a given dimensionless scale $\tau$, can be expressed as \cite{pope1993stationary} 
\begin{equation}
p(\Delta s)= \frac{N}{q_o(\Delta s)} \exp{ \int_0^{\Delta s} \frac{r_o(\Delta s')}{q_o(\Delta s')} d \Delta s' }.
\label{PopeChing_eq}
\end{equation}
This expression is exact when $\Delta s$ are realizations of a stationary stochastic process $S(t)$ under the condition $p(\Delta s) \to 0$ as $\Delta s \to \infty$. Here $q_o(\Delta s)= \langle \dot {S}^2 \vert \Delta s \rangle / \langle \dot {S}^2 \rangle $ and $r_o(\Delta s)= \langle \ddot {S} \vert \Delta s \rangle / \langle \dot {S}^2 \rangle $ are the normalized averages of the first and second order conditional derivatives of the process $S(t)$, and $N$ is a normalization constant. Eq. (\ref{PopeChing_eq}) generalizes previous results obtained by Sinai and Yakhot \cite{sinai1989limiting} and Ching \cite{ching1993probability} for the pdf of temperature fluctuations and their increments, where the term $r_o(\Delta s)$ was linear ($r_o(\Delta s)=-\Delta s$). Eq. (\ref{PopeChing_eq}), while derived for a twice-differentiable process, can be interpreted as the steady-state solution of a Fokker Planck equation with $p(\Delta s)$ vanishing at infinite boundaries, with drift and diffusion coefficient equal to  $r_0$ and $q_0$ respectively \cite{gardiner1985stochastic, porporato2011local}. 

Although Eq. (\ref{PopeChing_eq}) can be directly computed from an observed time series, the estimation of the conditional derivatives in $q_o(\Delta s)$ and $r_o(\Delta s)$ becomes inevitably uncertain as $\Delta s$ approaches the tails of the pdf. However, a number of parametric distributions commonly used in statistical mechanics arise as particular cases of Eq. (\ref{PopeChing_eq}) when $r_o(\Delta s)=-\Delta s$, such as Gaussian ($q_o$ constant), power-laws ($q_o(\Delta s) \sim \Delta s^2$) and stretched exponentials ($q_o(\Delta s) \sim \Delta s^a, 0<a<2$). To facilitate estimation and comparisons with data, two different parametric models for the tails of Eq. (\ref{PopeChing_eq}) are here adopted: a Stretched Exponential (SE) and a q-Gaussian distribution (QG). The first arises from multiplicative processes of normal-distributed random variates \cite{frisch1997extreme}, while the second maximizes a generalized measure of information entropy proposed by Tsallis \cite{tsallis1988possible, tsallis1995statistical,shi2005assessing}. While QG does not have a clear physical basis in the context of turbulent flows\cite{gotoh2004turbulence}, it has been widely used in the analysis of turbulence simulations and data \cite{Ramos2001non, arimitsu2002tsallis, bolzan2002analysis, katul2006analysis}. We employ these two models to infer tail behavior as well as to test the independence of our findings from the particular parametric distribution used to characterize $p(\Delta s)$. The QG and SE pdfs are given as

\begin{equation}
p_{QG}(\Delta s)=N(q) \cdot \left( 1+\left(q-1\right)\frac{\Delta s^2}{2 \psi^2} \right)^{\frac{1}{1-q}},
\label{QG}
\end{equation}
\begin{equation}
p_{SE}( \Delta s)= \frac{\eta}{\lambda} \left( \frac{ \Delta s }{\lambda} \right)^{\eta-1} \cdot \exp{ \left( \frac{ \Delta s }{\lambda} \right)^\eta}.
\label{SE}
\end{equation}

Both pdf models have two degrees of freedom corresponding to a scale ($\psi$,$\lambda$) and shape ($\eta,q$) parameter. We adopt the (symmetric) QG model and the SE fitted separately to right and left tails of $p(\Delta T)$.

\subsection{Probabilistic description of asymmetry and irreversibility across scales}

The presence of ramp-cliff structures has been conjectured to result in non-local interactions of different size eddies within the inertial subrange \cite{warhaft2000passive}.  This non-locality affects both even and odd moments of higher order. A statistical framework to investigate the effects of ramps on the asymmetric nature of velocity and scalar increments for different atmospheric stability classes is now discussed. Sharp edges associated with cliffs might directly inject scalar variance at much smaller scales and thus alter the magnitude and sign of odd order moments within the inertial range (depending on $z/L$). The presence of asymmetry has been object of investigations based on odd-ordered structure functions \cite{warhaft2000passive} or multipoint correlators \cite{mydlarski1998structures}. In particular, a simple measure for the persistence of asymmetry at small scales is the skewness of the scalar increments $S^3_T= \langle \Delta T(\tau)^3 \rangle / \langle \Delta T(\tau)^2 \rangle^{3/2}$. The structure skewness of air temperature has been found to scale as $Re_\lambda = \sigma_u \lambda /\nu$ (where  $\lambda$ is the Taylor microscale and $\sigma_u$ is the root mean square of the longitudinal velocity fluctuations) and thus for a boundary layer $S_3^T \sim Re_*^{1/2}$. However, for large values of $Re_\lambda$ experimental evidence suggests that $S_3^T$ tends to plateau and become independent of $Re_\lambda$ \citep{sreenivasan1991local,warhaft2000passive}.

A further signature of ramp-cliff structures is that increments $\Delta T(\tau)$ may exhibit a time directional (or 'irreversible') behavior. Time reversibility implies that the trajectories of a stationary process $\Theta_t$ exhibit the same statistical properties when considered forward or backward in time. In particular, for a reversible time series the n-points joint pdf of $(\Theta_1, \Theta_2, ...\Theta_n)$ is equal to the joint pdf of the reversed sequence $(\Theta_n, \Theta_{n-1}, ...\Theta_1)$ for every $n$. While testing this general definition of reversibility would require perfect knowledge of the phase space trajectories, a weaker definition is the so called lag-reversibility. This condition only requires the two-points pdfs to be equal: $f_{\Theta_t,\Theta_{t+\tau}}(\Theta_1,\Theta_2)=f_{\Theta_{t+\tau},\Theta_t}(\Theta_2,\Theta_1)$.  While this definition is less general, it still provides a necessary condition for testing time reversibility.  Moreover, it is consistent with the traditional descriptions of turbulence that are primarily based on two-point statistics. Lag reversibility implies that \cite{lawrance1991directionality}
\begin{equation}
R_{\tau}=\rho_c(\Theta_t^2,\Theta_{t+\tau})-\rho_c(\Theta_t,\Theta^2_{t+\tau})=0.
\label{corr_irr}
\end{equation}
where $\rho_c$ denotes a correlation coefficient. This condition can be directly tested across different $\tau$ and $\zeta$ using a conventional correlation analysis. 

A second test for reversibility of scalar trajectories is here performed based on the Kullback-Leibner measure, a form of relative entropy that determines the average distance between the entire pdf of forward and backward trajectories \cite{cover2012elements,porporato2007irreversibility,porporato2011local}. Again, the analysis here is restricted to the inspection of lag-reversibility $(n=2)$ across scales $\tau$. In such a restricted form, this measure reduces to

\begin{equation}
\langle Z_{\tau} \rangle = \int_{\Omega_{\Theta} }   \int_{\Omega_{\Theta'_{\tau} }} p(\Theta'_{\tau} |\Theta) p(\Theta) \log \frac{p(\Theta'_{\tau}|\Theta)}{{p}(- \Theta'_{\tau}|\Theta)} d \Theta'_{\tau}   d\Theta,
\label{rel_entr1}
\end{equation}

\noindent
where $ \Theta'_{\tau} = \Delta \Theta(\tau)/\tau $, and the domains of integration $\Omega_{\Theta} $ and $\Omega_{\Theta'_{\tau}} $ correspond to the populations of the random variables $\Theta$ and $\Theta'_{\tau}$ respectively. Eq. (\ref{rel_entr1}) determines, at each dimensionless scale $\tau$, the average distance between the probability of the transition $\Delta \Theta(\tau)$ and its inverse, at every given level $\Theta$. 

A statistical mechanics interpretation of Eq. (\ref{rel_entr1}) would imply that for a system in non-equilibrium steady state, the \emph{Fluctuation Theorem} must hold so that
 \begin{equation}
 \log{\frac{p(-Z_{\tau})}{p(Z_{\tau})}}=-Z_{\tau}
\label{FT}
\end{equation}
for the variable $Z_{\tau}$ computed at some level $\Theta$
 \begin{equation}
 Z_{\tau}(\Theta)=  \log \frac{p(\Theta'_{\tau}|\Theta)}{{p}(-\Theta'_{\tau}|\Theta)}.
\label{zFT}
\end{equation}
Note here the usage of conditional probabilities instead of their unconditional forms employed in recent flight-crash studies of Lagrangian fluid particles \cite{xu2014flight} that also made use of Fluctuation Theorem and time-reversibility. Eq. (\ref{rel_entr1}) has been shown to have general validity \cite{porporato2007irreversibility} independent of the underlying dynamics or statistical-mechanics interpretations, when considering conditional statistics. 
 
\section{Data and Methods}

The three velocity components and air temperature measurements were sampled at 56 Hz using an ultra-sonic anemometer positioned at $z=$5.2 m above a grass-covered surface at the Blackwood Division of the Duke Forest, near Durham, North Carolina, USA. The anemometer samples the air velocity in three non-orthogonal directions by transmitting sonic waves in opposite directions and measuring their travel times along a fixed 0.15 m path length. Temperature fluctuations are then computed from measured fluctuations in the speed of sound assuming air is an ideal gas. The non-orthogonal sonic anemometer design used here has proven to be the most effective at reducing flow distortions induced by the presence of the instrument.

The experiment resulted in 123 runs, each run having a duration of 19.5 minutes (65536 data points at 56$Hz$), covering a range of different atmospheric stability conditions\cite{katul1997energy}. The presence of a stable stratification is known to produce distortions on the spectral properties of turbulence at scales commensurate with (and larger than) the Dougherty-Ozmidov length scale \cite{rorai2015stably}. We investigated this issue (see the Appendix for more details) finding that stable stratification effects are only relevant at scales larger than the integral scale $I_w$ considered here and not in the inertial range.

\begin{figure}[ht]
\begin{center}
\centerline{\includegraphics[scale=0.8]{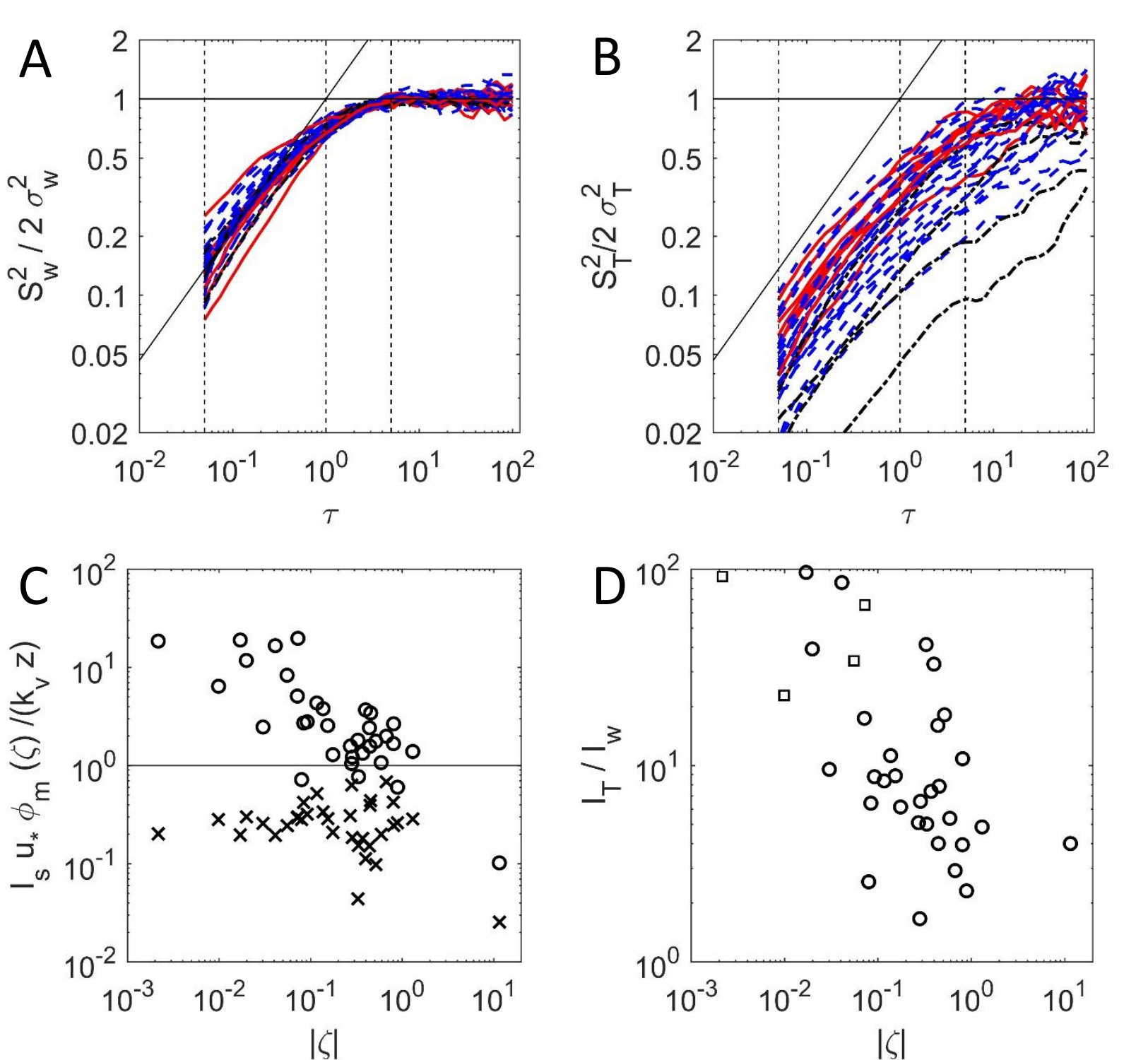}}
\caption{  In the upper panels, the normalized second order structure functions for vertical velocity (A) and temperature (B) are shown for runs that are weakly unstable (blue dashed lines), strongly unstable (red lines), and stable (black dash-dot lines). Black lines indicate the value 1 and the 2/3 power law for reference; vertical dashed lines correspond to the dimensionless scales $\tau=0.05$ (smallest scale not impacted by instrument path length), $\tau=1$ (integral scale of the flow), and $\tau=5$ (typical scale larger than $I_w$, while small enough not to be impacted by statistical convergence issues in structure functions calculations). Lower panels illustrate (C) the integral scales of the flow for $s=T$ (circles) and $s=w$ (crosses) as a function of the stability parameter $|\zeta|$, and (D) their ratio $I_T$ to $I_w$ again as a function of $|\zeta|$, where stable runs ($\zeta>0$) are indicated by black squares. }
\label{Int_scales}
\end{center}
\end{figure}

The assumption of stationarity is necessary so as to (i) decompose the flow variables into a mean and fluctuating part, (ii) adopt Eqs. (\ref{PopeChing_eq}) and (\ref{rel_entr1}) so as to describe intermittency and time irreversibility respectively, and (iii) compute the integral scales needed in delineating the transition from production to inertial. To test the dataset for stationarity, we employ the second order structure functions of velocity components ($u,w$) and air temperature $T$. Runs were included only if the slope of  $S^2_s=\langle \left[ s(t+\tau)-s(t) \right]^2 \rangle $ for time delays larger than about 9 minutes (30000 sample points) was smaller than a fixed value ($0.01$). Only 34 runs were retained based on this strict stationarity criterion. Their corresponding second order structure functions for $w$ and $T$ are featured in Fig. \ref{Int_scales}. As expected, structure functions exhibit an approximate $2/3$ scaling at fine scales and transition to a constant value as the autocorrelation weakens at large separation distances.

As earlier noted, the most restrictive (i.e. smallest) integral time scale is $I_w$ associated with the vertical velocity $w$ due to ground effects. We assume that this time scale characterizes the transition from production to inertial ranges for all three flow variables $u,w,T$. Eq (\ref{Iw}) is here evaluated by integrating $\rho_w (\tau)$ up to the first zero crossing so as to avoid the effects of low frequency oscillations.  Figure \ref{Int_scales} illustrates the integral time scales of $w$ and $T$ as a function of $\zeta$, where the aforementioned integral time scales are normalized by the mean vorticity time scale $dU/dz=\phi_m(\zeta) u_* (\kappa_v z)^{-1}$.  It is clear that such normalized $I_w$ is approximately constant across stability regimes and suggests $I_w$ to be proportional to the duration of vortices most efficient at transporting momentum to the ground for all $\zeta$.  Conversely, the temperature integral time scale is much longer than $I_w$ for near-neutral conditions and only approaches $I_w$ for strongly unstable conditions.  

A known limitation of sonic anemometry is the presence of distortions at high frequencies due to instrument path-averaging. For this reason, the smallest time scale considered in the analysis is $0.05 \cdot I_w$, which corresponds to a minimum travel path of $30cm$ (or twice the sonic anemometer path length). Taylor's frozen turbulence hypothesis \cite{taylor1938spectrum} ($r=-\overline{U}t$) was employed to convert values of $\tau$ to separation distances $r$ within the inertial subrange even though the turbulent intensity $\sigma_u /U$ is not small as shown in Table \ref{tab:1}. For this reason, we adopt the dimensionless lag $\tau$ for analysis and presentation. The $\tau$ can be interpreted as temporal or spatial noting that distortions due to the use of Taylor's hypothesis impact similarly the numerator and denominator.

\begin{figure}[ht]
\begin{center}
\centerline{\includegraphics[scale=0.8]{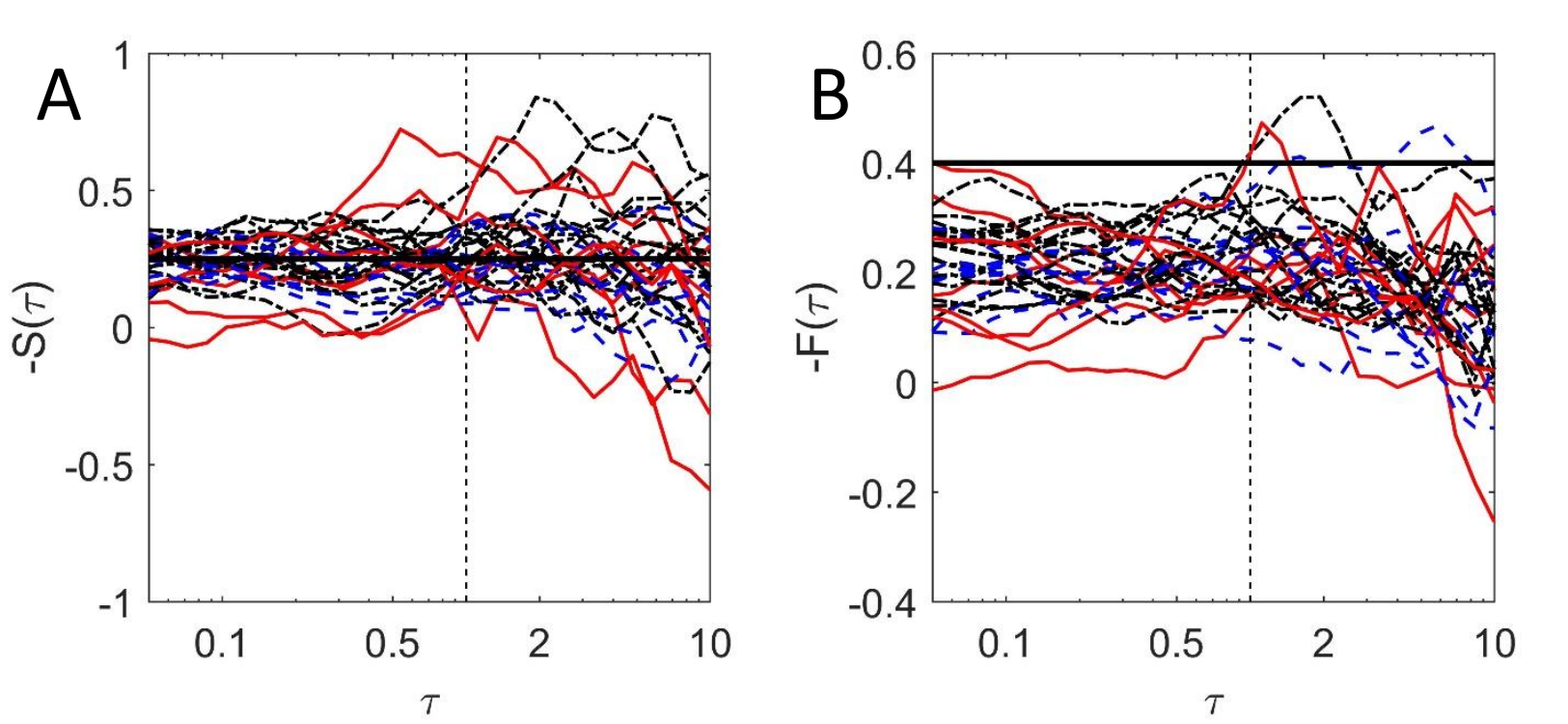}}
\caption{Normalized third order structure functions $S(\tau)$ and $F(\tau)$ at the crossover from inertial to production scales. Vertical dashed line indicates the integral time scales, horizontal lines show the constant values $0.25$ (A) and $0.4$ (B). Results are shown for weakly unstable runs (blue dashed lines), strongly unstable (red lines), and stable runs (black dash-dot lines).}\label{scaling_FrSr}
\end{center}
\end{figure}

To compare the data sets here with laboratory studies, a number of statistics were computed and presented. The validity of Obukhov's constant skewness hypothesis was tested for $u$ in Figure \ref{scaling_FrSr}, which reports the values of the third order structure functions Eqs. (\ref{Sr}) and (\ref{Fr}) evaluated at the onset of the inertial subrange delineated by the $w$ time series. Both are approximately constant for scales smaller than $I_w$. While comparison with experiments shows good agreement for $S(\tau) \simeq -0.25$, $F(\tau)$ is systematically smaller than its anticipated value \cite{katul1997energy} ($-0.4$) for all $\zeta$.

\begin{figure}[ht]
\begin{center}
\centerline{\includegraphics[scale=0.8]{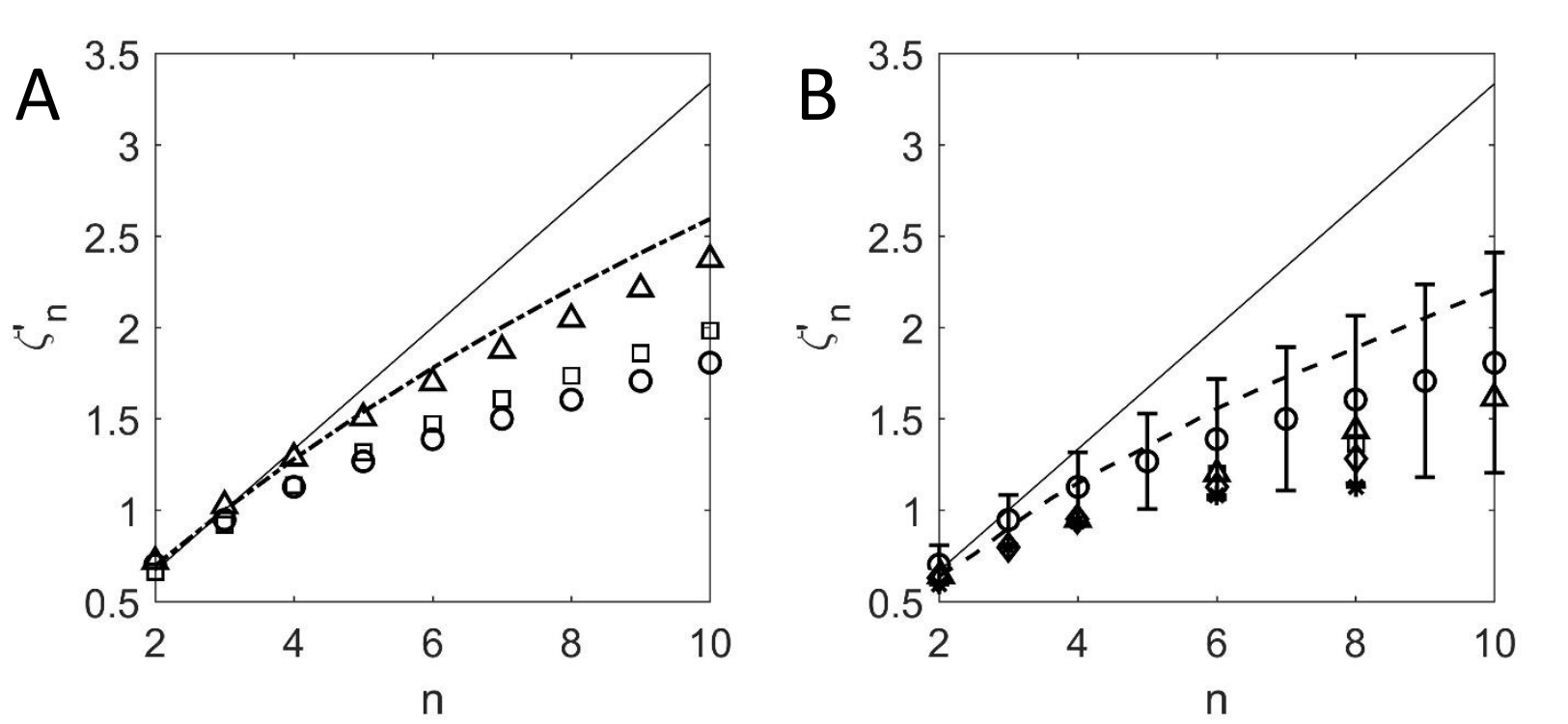}}
\caption{(A) Average values of the scaling exponents for longitudinal velocity $u$ (triangles), vertical velocity $w$ (squares), and temperature $T$ (circles). Black continuous line and dashed line show respectively the K41 and the She-leveque predictions for the longitudinal velocity structure functions. Exponents are computed from scales ranging between $\tau = 0.05$ and $\tau = 0.2 $. (B) Scaling exponents for temperature only; Mean and standard deviation values are computed over all the runs and are indicated by circles and vertical bars, respectively. Data from Mydlarsky and Warhaft (1990) \cite{mydlarski1998passive} (squares), Antonia et al. (1984) \cite{antonia1984temperature} (triangles), Meneveau et al. (1990) \cite{meneveau1990joint} (*) and Ruiz et al. (1996) (diamonds) \cite{ruiz1996scaling} are shown for comparison. The KOC scaling (black line) and results from the Kraichnan model (1994) \cite{kraichnan1994anomalous} (dashed line) as reported in \cite{warhaft2000passive} are also presented for reference.  }\label{scaling_exponents}
\end{center}
\end{figure}

Inspection of scaling exponents $\zeta'_{n}$ in Eq. (\ref{scaling_anomaly}) for $u,w,T$ confirms that K41 predictions significantly overestimate scaling exponents for structure functions of order higher than 2, as shown in figure \ref{scaling_exponents}(A). The scaling exponents obtained for the scalar $T$ show reasonable agreement with previous experimental results (Fig. \ref{scaling_exponents}(B)), with values systematically lower than predicted by the Kraichnan model in the limiting case of time-uncorrelated velocity field \cite{kraichnan1994anomalous}.

For every run, $\zeta$ was computed using Eq. (\ref{TKE_Production2}) and then employed to classify the ASL stability condition. Most of the runs in the dataset are unstable with a wide range of $\left| \zeta \right|$, while only 4 runs are characterized by $\zeta>0$. To ensure a balanced statistical design, two stability classes are selected with the same number of runs (8) in each class: strongly unstable ($\left| \zeta \right| > 0.5$) and near neutral runs ($\left| \zeta \right| <0.072$). A summary of the bulk flow properties for these runs are featured in Table (\ref{tab:1}). 

In the analysis, each flow variable $s$ ($s=u,w,T$) is normalized to zero-mean and unit-variance (labeled as $s_n$). Then, at scale $\tau$, a time series of $\Delta s (\tau) = s_n (t+\tau) - s_n (t)$ is constructed and again normalized to have unit variance. 

\begin{figure}[ht]
\begin{center}
\centerline{\includegraphics[scale=0.6]{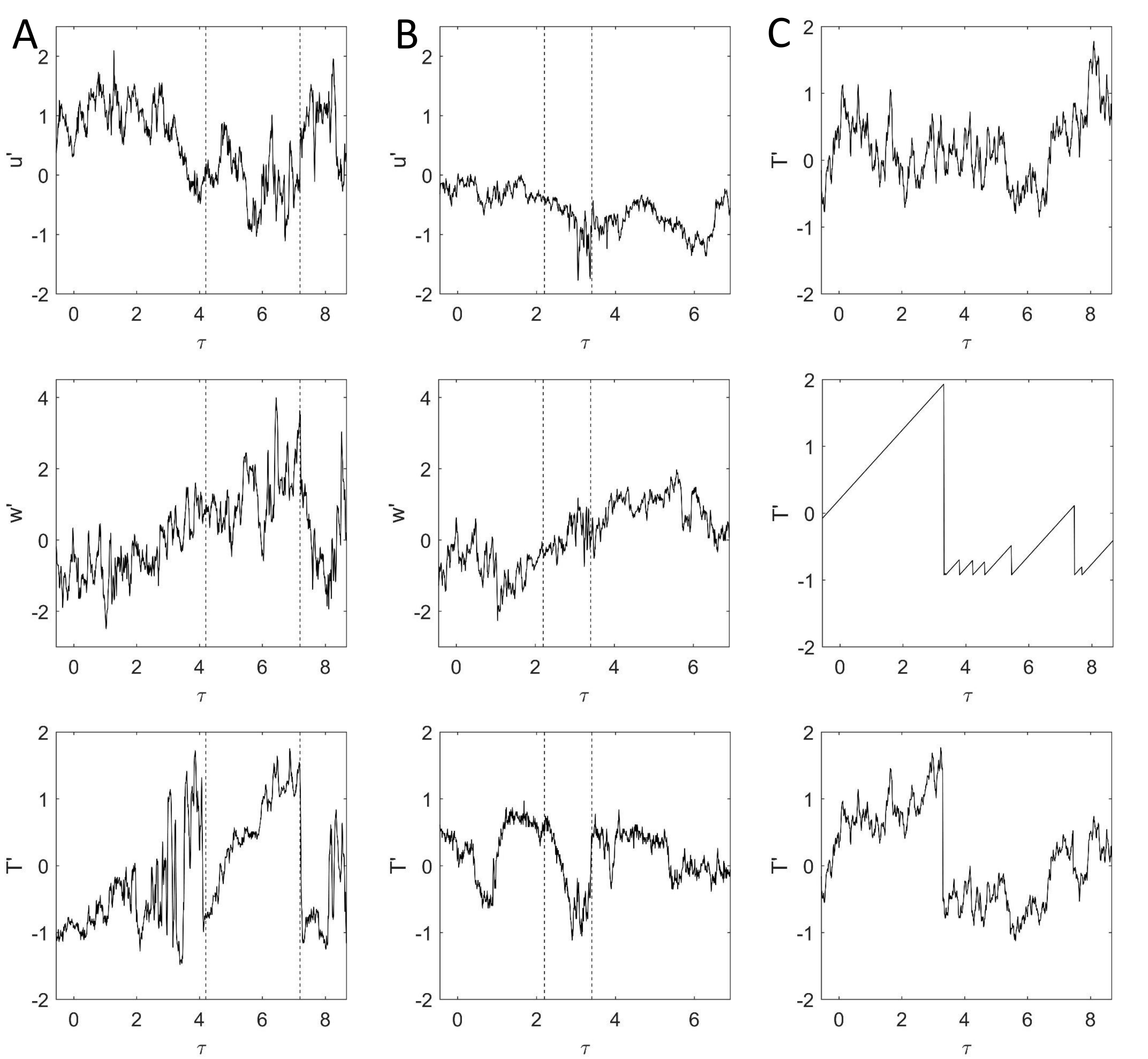}}
\caption{Sequences of velocity and temperature fluctuations extracted from a strongly unstable run (run 8, $\zeta = -0.52$, $I_w=  1.74 s$, column A) and a stable/near neutral one (run 34, $\zeta = 0.07 $, $I_w= 2.18 s$ column B). The presence of ramps and inverted-ramp like structures respectively is marked by dashed vertical lines. Column (C) illustrates a phase-randomized sequence obtained from run 34 (top), a series of synthetic ramps with durations exponentially distributed with mean 2$I_w$ (middle) and the surrogate time series obtained merging the above sawtooth pattern with the phase randomized time series (bottom), where the relative weight of the ramps $\alpha$ was set equal to $0.5$.}\label{timeseries}
\end{center}
\end{figure}

For illustration purposes, Fig. \ref{timeseries} shows sequences of fluctuations $u',w',T'$ extracted from runs in unstable and stable atmospheric regimes. In the first case, temperature fluctuations clearly exhibit ramp-cliff structures occurring with time scales larger than $I_w$. In the stable/near neutral case, large scale scalar structure are still present even though their structure is qualitatively different from the unstable case, and may include inverted ramp structures as in Fig. \ref{timeseries}(B) when $\overline{w'T'}<0$.

To test the effects of these coherent structures on inertial subrange statistics, and in particular to isolate the effect of temperature ramps on intermittency and asymmetry, synthetic time series are used and are constructed as follows. First, a phase-randomization of the original temperature records \cite{prichard1994generating} is performed by preserving the amplitudes of the Fourier coefficients while destroying coherent patterns encoded in the phase angle. A synthetic sawtooth time series is then superimposed on the time series obtained by phase-randomization. Here a coefficient $\alpha$ measures the relative weight of the ramps with respect to the phase-randomized sequence. This combination yields realizations of a renewal process (see Fig. \ref{timeseries}(C) for a representative example with $\alpha=0.5$) that is unconnected with Navier-Stokes scalar turbulence, but mimics sweep-ejection dynamics\cite{katul2006analysis}. Synthetic ramps are here generated with exponentially distributed durations and with a mean duration set to a multiple of the integral time scale ($2\cdot I_w$ in Figure \ref{timeseries}(C)). The resulting time series is again normalized to have zero mean and unit variance. 

Eq. (\ref{rel_entr1}) was computed by integrating the relative entropy over the joint frequency distribution of normalized temperature fluctuations and their increments at each scale $\tau$. We use a coarse binning for estimating the joint pdf $ p(T'(\tau),T)$ and assume \cite{porporato2007irreversibility} that only finite probability ratios contribute to $\langle Z_{\tau} \rangle$. To check the consistency of this approach, calculations of Eq. (\ref{rel_entr1}) were repeated using a phase space reconstruction technique based on embedding sequences ($T_t, T_{t+\tau}$) with delay time $\tau$ and embedding dimension 2, which confirmed the validity of this approach (results not shown).

\begin{longtable}{c c c c c c c c c c c}
\caption{Bulk flow properties for the runs in our dataset. The table reports the atmospheric stability parameter $\zeta$, the Obukhov length $L \quad [m]$, the sensible heat flux $H = \rho C_p \overline{w'T'} \quad [W m^{-2}]$ (where $\rho$ is the mean air density and $C_p$ is the specific heat capacity of dry air at constant pressure), the mean air temperature $T \quad [^{\circ} C]$ and mean velocity $U \quad [m/s]$, and the integral time scale for $w$ $[s]$, the turbulent intensity $\sigma_u /U$, the temperature standard deviation $\sigma_T \quad [^{\circ} C]$, and vertical velocity standard deviation $\sigma_w \quad [m/s]$.}
\label{tab:1}\\

\hline
Run                & $\zeta$ & $L$ & $H  $ & ${T}$ & ${U} $  & $I_w$  & $ \sigma_u /U $ & $u^*$ & $\sigma_T $ & $\sigma_w $\\
\hline
\endfirsthead

\hline
Run                & $\zeta$ & $L$ & $H  $ & ${T}$ & ${U} $  & $I_w$  & $ \sigma_u /U $ & $u^*$ & $\sigma_T $ & $\sigma_w $\\
\hline
\endhead

1   & -11.56 & -0.4   & 93.2  & 33.9  & 2.1   & 2.62  & 0.44 & 0.08  & 0.48 & 0.40 \\
2   & -1.31  & -4.0   & 121.6 & 26.9  & 1.0   & 7.58  & 0.72 & 0.17  & 0.54 & 0.30 \\
3   & -0.89  & -5.8   & 73.1  & 27.8  & 0.5   & 6.62  & 0.91 & 0.16  & 0.37 & 0.30 \\
4  & -0.81  & -6.4   & 79.9  & 32.7  & 0.7   & 5.75  & 1.05 & 0.17  & 0.61 & 0.29 \\
5   & -0.80  & -6.5   & 138.1 & 27.4  & 0.8   & 8.18  & 0.48 & 0.21  & 0.57 & 0.31 \\
6  & -0.67  & -7.7   & 149.8 & 31.4  & 0.9   & 11.64 & 1.04 & 0.23  & 0.63 & 0.38 \\
7  & -0.59  & -8.8   & 118.1 & 34.8  & 1.5   & 3.43  & 0.71 & 0.22  & 0.58 & 0.34 \\
8  & -0.52  & -10.0  & 85.4  & 32.5  & 2.1   & 1.74  & 0.37 & 0.21  & 0.44 & 0.37 \\
\hline
9  & -0.45  & -11.5  & 78.6  & 31.7  & 1.1   & 7.44  & 0.61 & 0.21  & 0.43 & 0.30 \\
10  & -0.44  & -11.7  & 110.7 & 31.9  & 1.2   & 5.89  & 0.65 & 0.24  & 0.49 & 0.37 \\
11   & -0.44  & -11.8  & 39.4  & 34.4  & 1.3   & 3.19  & 0.45 & 0.17  & 0.32 & 0.29 \\
12   & -0.40  & -13.0  & 36.6  & 34.1  & 1.7   & 2.30  & 0.39 & 0.17  & 0.37 & 0.28 \\
13   & -0.37  & -14.0  & 65.1  & 25.2  & 1.6   & 2.91  & 0.39 & 0.21  & 0.35 & 0.27 \\
14  & -0.33  & -15.6  & 48.0  & 28.9  & 1.4   & 2.58  & 0.41 & 0.20  & 0.27 & 0.30 \\
15  & -0.33  & -15.8  & 4.8   & 33.4  & 1.6   & 1.59  & 0.35 & 0.09  & 0.09 & 0.23 \\
16  & -0.29  & -18.2  & 115.2 & 32.1  & 2.7   & 2.16  & 0.37 & 0.28  & 0.44 & 0.47 \\
17   & -0.28  & -18.5  & 136.2 & 29.2  & 0.9   & 6.88  & 1.11 & 0.30  & 0.56 & 0.37 \\
18  & -0.27  & -19.1  & 108.6 & 30.5  & 1.7   & 3.56  & 0.62 & 0.28  & 0.54 & 0.34 \\
19  & -0.17  & -29.7  & 70.5  & 29.5  & 2.6   & 2.22  & 0.29 & 0.28  & 0.36 & 0.42 \\
20  & -0.15  & -33.8  & 63.2  & 32.9  & 2.2   & 2.97  & 0.39 & 0.28  & 0.36 & 0.40 \\
21  & -0.14  & -37.9  & 30.9  & 34.2  & 1.6   & 4.17  & 0.49 & 0.23  & 0.34 & 0.32 \\
22  & -0.12  & -44.4  & 118.6 & 31.0  & 2.6   & 3.78  & 0.42 & 0.38  & 0.49 & 0.42 \\
23  & -0.09  & -56.5  & 26.7  & 33.9  & 1.9   & 3.39  & 0.31 & 0.25  & 0.15 & 0.31 \\
24  & -0.08  & -61.7  & 49.7  & 31.7  & 2.0   & 3.50  & 0.41 & 0.31  & 0.27 & 0.39 \\
25   & -0.08  & -65.1  & 17.6  & 34.0  & 2.2   & 3.22  & 0.29 & 0.23  & 0.13 & 0.31 \\
\hline
26  & -0.07  & -72.5  & 28.8  & 31.5  & 1.8   & 2.71  & 0.41 & 0.28  & 0.29 & 0.30 \\
27  & -0.04  & -126.2 & 45.1  & 31.0  & 4.3   & 1.21  & 0.33 & 0.39  & 0.35 & 0.71 \\
28  & -0.03  & -171.8 & 3.9   & 31.3  & 1.7   & 3.18  & 0.39 & 0.19  & 0.15 & 0.30 \\
29  & -0.02  & -261.4 & 46.1  & 31.2  & 3.8   & 1.37  & 0.39 & 0.50  & 0.23 & 0.72 \\
30  & -0.02  & -304.3 & 47.1  & 29.4  & 5.0   & 0.84  & 0.31 & 0.53  & 0.21 & 0.80 \\
\hline
31  & 0.002   & 2397.4 & -0.4  & 31.2  & 1.9   & 1.94  & 0.44 & 0.22  & 0.69 & 0.32 \\
32  & 0.01   & 525.5  & -1.3  & 32.9  & 0.9   & 3.00  & 0.51 & 0.19  & 0.18 & 0.23 \\
33  & 0.05   & 93.8   & -20.7 & 29.8  & 2.6   & 1.52  & 0.30 & 0.27  & 0.23 & 0.39 \\
34  & 0.07   & 71.4   & -14.2 & 30.4  & 1.9   & 2.18  & 0.37 & 0.22  & 0.25 & 0.28 \\
\hline
\end{longtable}

\section*{Results}
The main questions to be addressed require determination of the scale-wise evolution of (i) the probability of extreme scalar concentration excursions and concomitant intermittency buildup, and (ii) symmetry and time reversibility. These two questions are explored using the data sets here for stable, near neutral and unstable ASL runs.

\subsection{Probabilistic description of intermittency across scales}
\begin{figure}[ht]
\begin{center}
\centerline{\includegraphics[scale=0.6]{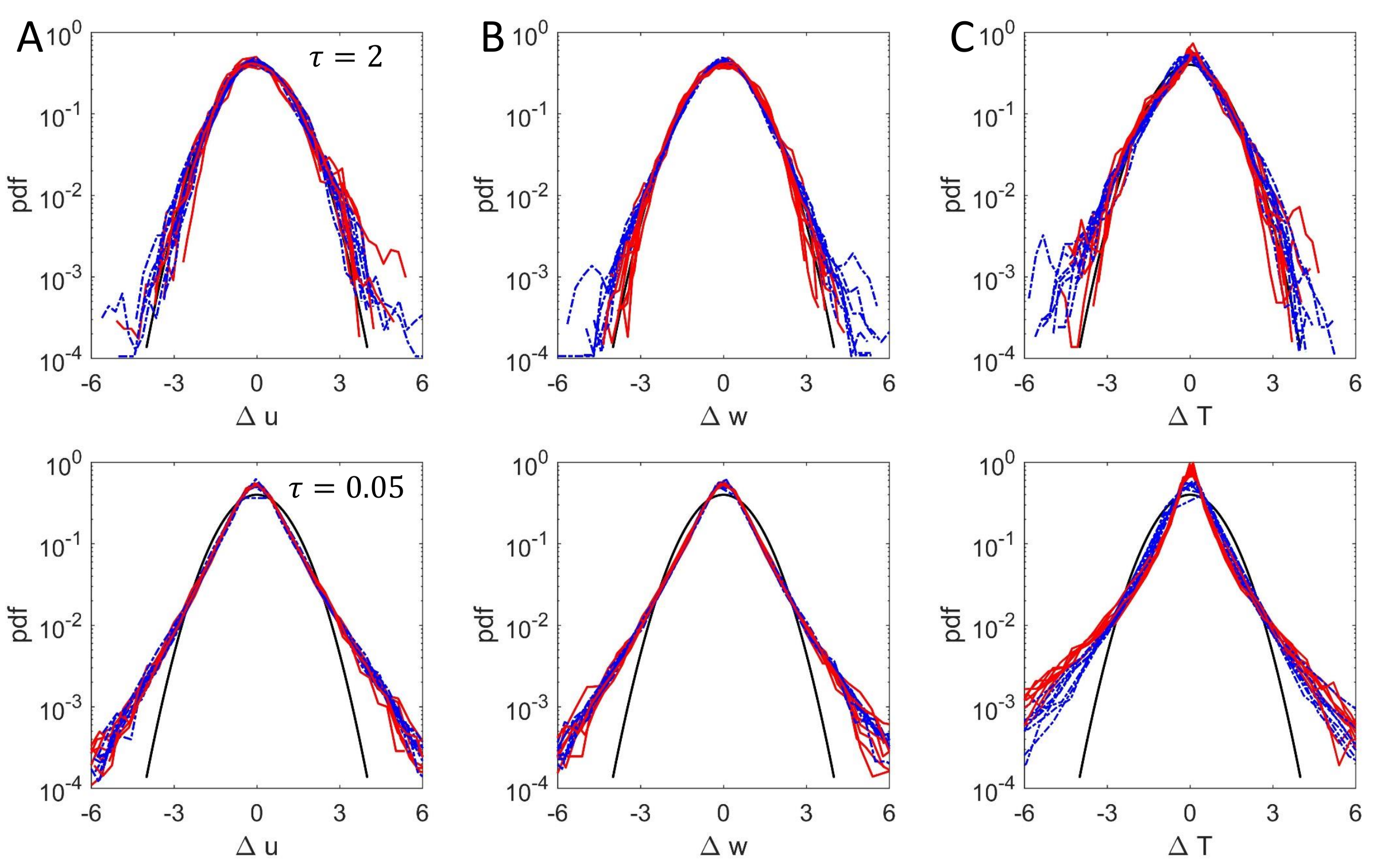}}
\caption{Normalized probability density functions observed for increments of longitudinal velocity (A), vertical velocity (B) and air temperature (C) at large scales ($\tau=2$, top panels) and small scales ($\tau = 0.05$, lower panels). The figure includes data from runs in the strongly unstable class ($\zeta < -0.5$, shown in red), and near neutral class ($| \zeta |< 0.072$, blue). Black lines show the standard Gaussian distribution for reference.}\label{pdfs}
\end{center}
\end{figure}

The empirical pdfs of velocity and air temperature increments ($\Delta s=\Delta u,\Delta w, \Delta T$) for runs in the near-neutral ($\left| \zeta \right| <0.072 $) and strongly unstable ($\zeta<-0.5$) classes (Fig. \ref{pdfs}) show clear transitions from a quasi-Gaussian regime at large lags ($\tau=2$ in figure) to distributions with sharper peaks and longer tails at scales well within the inertial subrange ($\tau=0.05$). This behavior has been documented for a wide range of turbulent flows \cite{meneveau1991analysis} and is associated with the build up of intermittency \cite{kolmogorov1962refinement} due to self-amplification inertial dynamics \cite{li2005origin}. 

\begin{figure}[ht]
\begin{center}
\centerline{\includegraphics[scale=0.8]{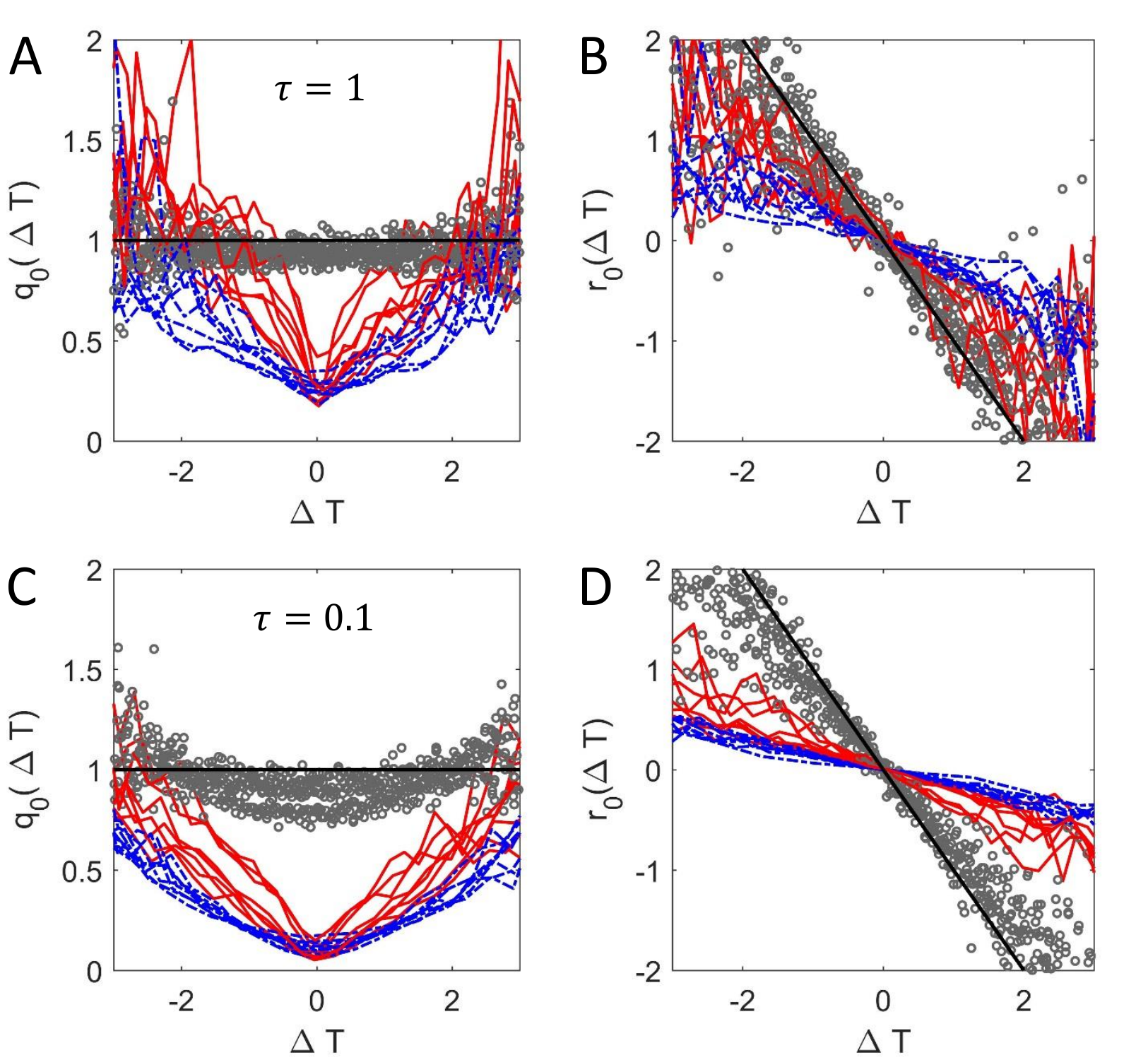}}
\caption{Functions $q_0(\Delta T)$ and $r_0(\Delta T)$ estimated from the conditional derivatives of the original temperature time series, for the two classes of strongly unstable (red lines) and near neutral runs (blue dashed lines). The same quantities are reported for phase-randomized surrogate time series for comparison (grey circles). Results are shown for the central body of the pdf (within $3 \sigma$ from the mean) for illustration purposes. Top panels (A,B) are computed for a lag equal to the integral time scale of the flow $\tau=1$, while the bottom panels (C,D) correspond to the smaller time lag $\tau = 0.1$.  Black lines $q_0=1$ and $r_0=-\Delta T$ correspond to the standard Gaussian distribution.}\label{PopeChing}
\end{center}
\end{figure}

The bulk of the pdf of temperature increments at any given scale can also be characterized by the coefficients of Eq. (\ref{PopeChing_eq}). Results show some differences between runs with differing $|\zeta|$ (Fig. \ref{PopeChing}). Namely, for runs in the strongly unstable class, $q_0$ exhibits a more pronounced peak around the origin and is characterized by larger asymmetry at the cross-over scale $\tau=1$ compared to their near-neutral counterparts (Fig. \ref{PopeChing}(A)). Moreover, the results here confirm that a choice of linear $r_0(\Delta T)$ and quadratic $q_0(\Delta T)$ appear reasonable for ASL flows. In the case of an unstable ASL, the term $r_0(\Delta T)$ remains linear, while inspection of $q_0(\Delta T)$ suggests that a dependence on $s$ with an exponent smaller than 2 might be more appropriate, corresponding to stretched exponential tails for $p( \Delta T)$ for small lags $\tau$ in unstable ASL flows. Comparison with the same data after run-by-run spectral phase randomization \cite{prichard1994generating} shows that the latter exhibits almost Gaussian behavior, confirming that the emergence of long tails at inertial scales is primarily a consequence of non linear structures in the original time series. 

\begin{figure}[ht]
\begin{center}
\centerline{\includegraphics[scale=0.6]{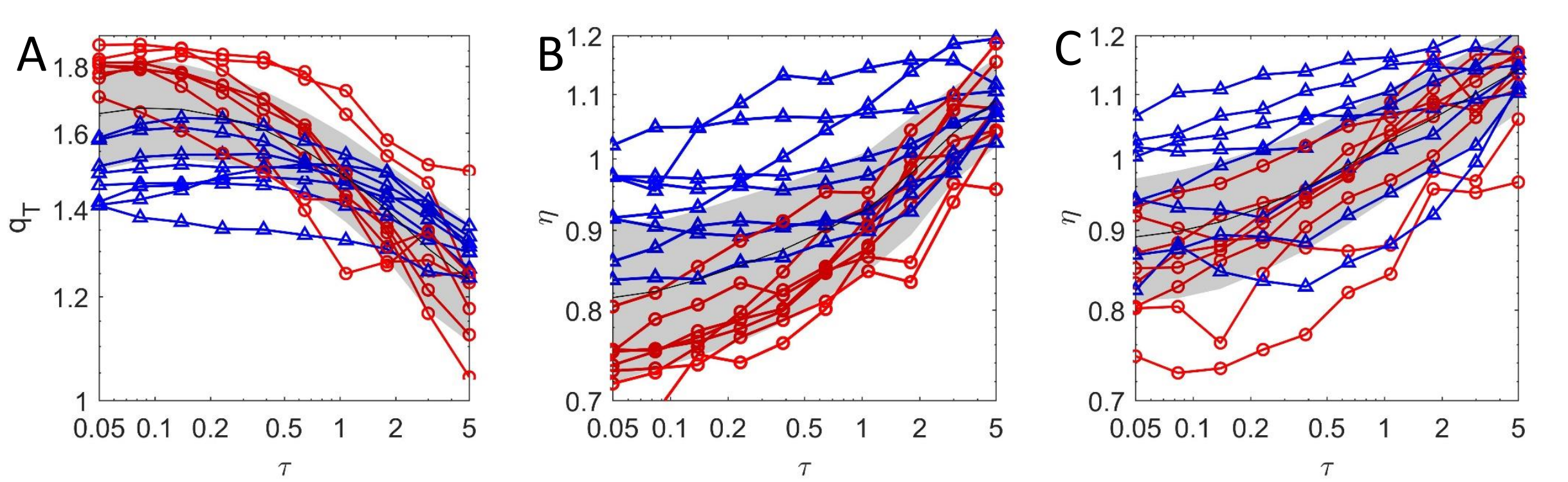}}
\caption{Evolution across scales $\tau$ of the q-Gaussian tail parameter $q$ (A), and of the stretched exponential shape parameter $\eta$ obtained from separate fit to the left (B) and right (C) tails of the distribution of temperature increments. Data from two stability classes are included: strongly unstable ($\zeta < -0.5$, red cirles) and near neutral conditions ($| \zeta |< 0.072$, blue triangles). Black lines and shaded areas indicate average values and standard deviations respectively computed over the entire dataset.}\label{scalewise_evol}
\end{center}
\end{figure}

\begin{figure}[ht]
\begin{center}
\centerline{\includegraphics[scale=0.6]{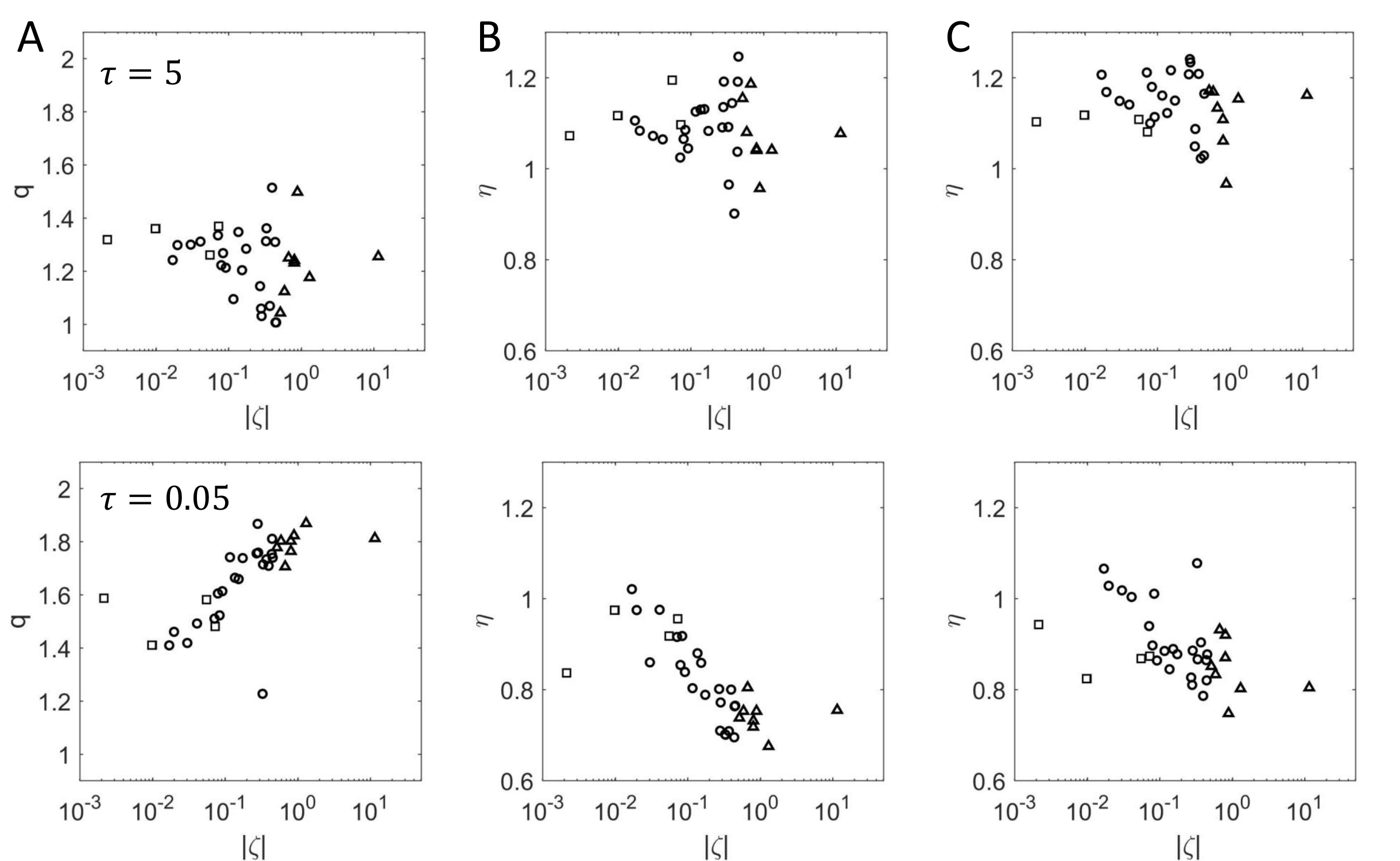}}
\caption{Tail parameters of the pdf of temperature increments across stability conditions $\zeta$. Results include the q-Gaussian tail parameter $q$ (column A) and the stretched exponential shape parameter $\eta$, obtained from fitting the left (column B) and right tail (column C) of the distribution $p(\Delta T)$. Values of $q$ and $\eta$ are reported for large scales ($\tau = 5$, upper panels) and small scales ($ \tau = 0.05$, lower panels). Triangles denote strongly unstable runs ($\zeta < -0.5$), squares denote stable runs ($\zeta > 0$) and circles refer to slightly unstable runs ($-0.5 <\zeta <0$).}\label{tail_param}
\end{center}
\end{figure}

The variation of the tail parameters \textbf{$\eta$} and $q$ with decreasing scale $\tau$ (Fig. \ref{scalewise_evol}) provides a robust measure of how the distributional tails of $p( \Delta T)$ evolve at the onset of the inertial range. For temperature differences, the rates of change across scales of both $\eta$ and $q$ appear to be dependent on the magnitude of the stability parameter $\zeta$. Consequently, while at large scales - where the pdf closely resembles a Gaussian -  neither \textbf{$\eta$} nor $q$ exhibit a significant dependence on $\zeta$, for scales well within the inertial subrange stability is clearly impacting the tail behavior of $\Delta T$ (Fig. \ref{tail_param}).

This evidence suggests that the observed intermittency is not only internal (i.e., not only due to variability in the instantaneous dissipation rate\cite{kuznetsov1992fine}) but is also directly impacted by the larger scale eddy motion that sense boundary conditions. In particular, when buoyancy generation is significant, the heat flux $\overline{w'T'}$ is connected with the sweep and sudden ejection of air parcels, corresponding with the sharp edges of the temperature ramps \cite{antonia1979temperature,katul2006analysis}. The resulting sawtooth behavior could be responsible for the injection of scalar variance at small scales (instead of a gradual cascade), acting in particular on the negative tail of the $\Delta T$ pdf, as evident from Fig. \ref{PopeChing}(A). On the other hand, the buildup of non-Gaussian statistics for velocity increments is not as impacted by the stability regime, and therefore the dominant effects are in this case primarily an effect of internal intermittency.

\subsection{Probabilistic description of asymmetry across scales}

\begin{figure}[ht]
\begin{center}
\centerline{\includegraphics[scale=0.8]{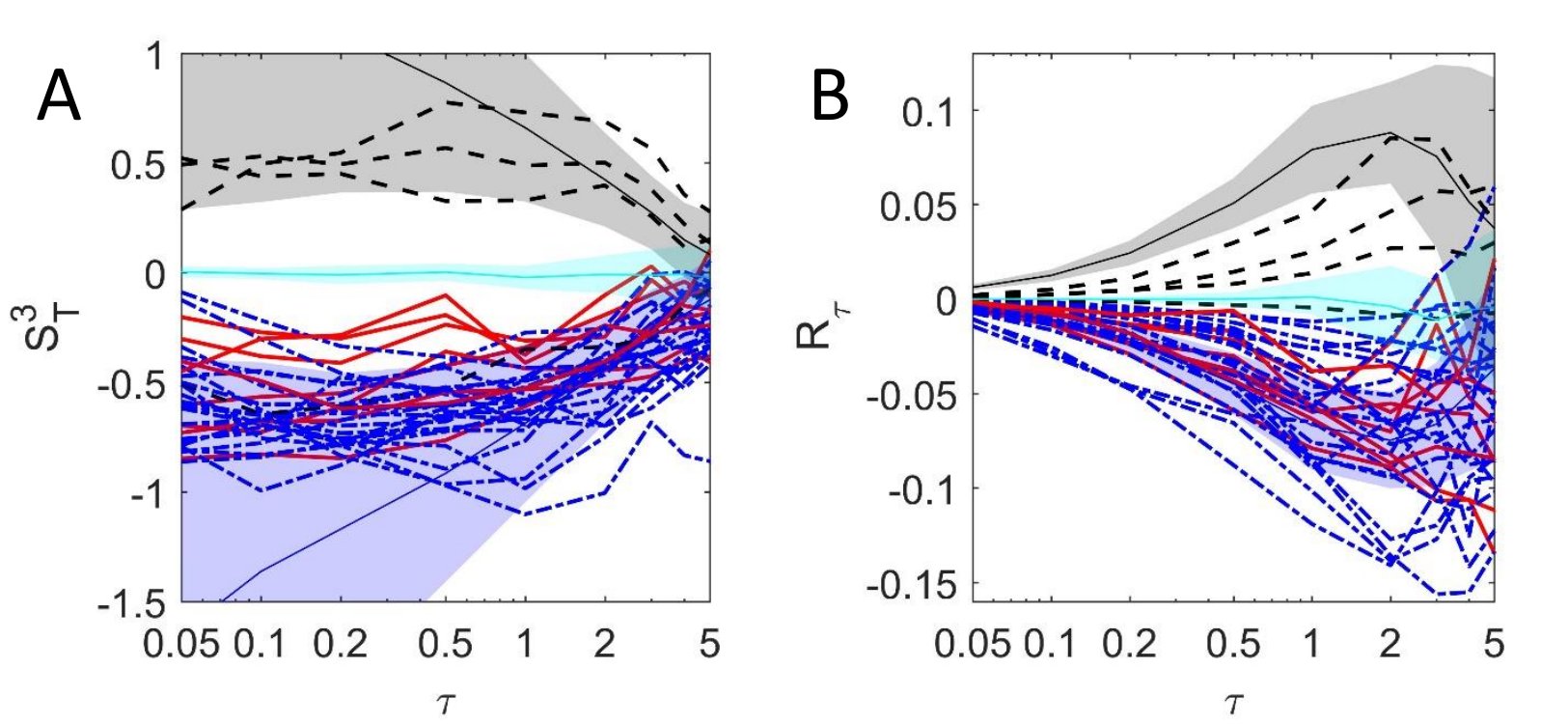}}
\caption{Measures of asymmetry $S^3_T$ (A) and time irreversibility $R_{\tau}$ (B) computed for temperature increments for scales varying from $\tau=0.05$ to $\tau=5$. The plots include stable runs (black dashed lines), weakly unstable runs (blue dash-dot lines) and strongly unstable runs (red lines). For reference, the same quantities are computed for phase-randomized time series (cyan), and sythetic time series with sawtooth positive (blue) and inverted ramps (black). Shaded regions correspond to the $1\sigma$-confidence intervals over 34 realizations of the surrogate time series. Relative weight and mean duration of the synthetic ramps were set to $\alpha = 0.4$ and 2$ I_w$ respectively.}\label{Irrev}
\end{center}
\end{figure}

The presence of a finite third order temperature structure function signifies that local isotropy is not fully attained in the range of scales explored here. The skewness $S^3_T$ exhibits a plateau for scales smaller than $I_w$ (Fig. \ref{Irrev}(A)) similar to previous measurements reported in grid turbulence forced by a mean temperature gradient \cite{mydlarski1998passive}. Moreover, $S^3_T$ levels off to positive values for $\zeta>0$, while it becomes negative for $\zeta<0$. This finding is consistent with the presence of ramp-like structures when $\zeta >0$ (mildly stable conditions) that are inverted when compared to their unstable counterparts. 

The findings here confirm that at the cross-over from production to inertial, imprints of ramp structures persists well into the inertial subrange. The consequence of these imprints on time-reversibility is now considered for temperature sequences. The irreversibility analysis detects strong irreversbility at large scales that slowly decreases at the onset of the inertial range (Fig. \ref{Irrev}). This finding is consistent with the idea that atmospheric stability determines a preferential direction for the large-scale scalar structures, which becomes progressively weaker at scales smaller than $\tau = 1$. Here the sign of the heat flux has a primary effect on the orientation of the ramps, as captured by $R_{\tau}$. Furthermore, phase randomization is shown to destroy much of this time irreversibility (Fig. \ref{Irrev}(B)) while the addition of synthetic ramps, either with positive or negative orientation, produces values of $R_{\tau}$ that closely resemble observations of stable and unstable ASL respectively. These synthetic experiments also recover the sign of the third order moment $S^3_T$ (Fig. \ref{Irrev}(A)) but not its magnitude at smaller scales. As one would expect, a sawtooth time series does not fully reproduce inertial scale scalar dynamics, even though it does clearly capture the effect of boundary conditions on scalar ramp-cliffs.

\begin{figure}[ht]
\begin{center}
\centerline{\includegraphics[scale=0.8]{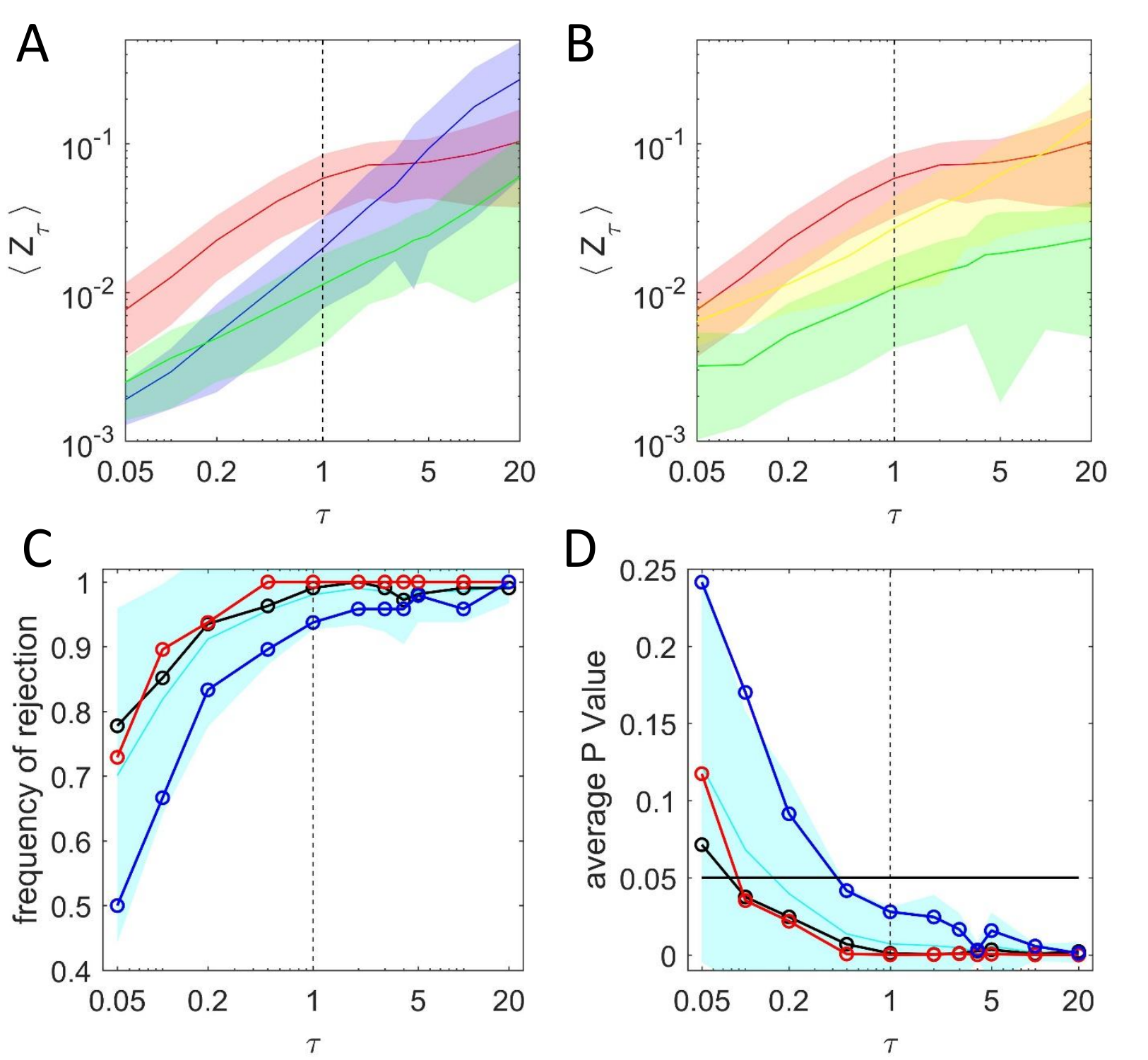}}
\caption{(A) Mean and standard deviation over 34 time series of $ \langle Z_{\tau} \rangle$ computed for scales varying from $\tau=0.05$ to $\tau=20$. Values of $ \langle Z_{\tau} \rangle$ are shown for original temperature records (red), and surrogate time series obtained by phase-randomization (green). For comparison, the same analysis is reported for fractional brownian motion with Hurst exponent $H=1/3$ (blue). (B) A comparison of $\langle Z_{\tau} \rangle$ for temperature (red), longitudinal velocity (yellow) and vertical velocity (green). The lower panel shows the Kolmogorov-Smirnov test average rejection rate (C) and average P-value (D) computed for all the temperature time series (cyan for mean value and 1$ \sigma$ confidence interval), and for different stability classes: strongly unstable runs ($ \zeta < -0.5$, red), near-neutral runs ($ \vert \zeta \vert < 0.072 $, blue) and intermediate values ( $ 0.072 < \vert \zeta \vert < 0.5$, black). KS test was performed at the 0.05 significance level, corresponding to the horizontal line in (D). The vertical dashed line marks the integral time scale $I_w$. }\label{shaded}
\end{center}
\end{figure}

The averaged relative entropy $\langle Z_{\tau} \rangle$, while insensitive to the ramp orientation, at every given level $T$ quantifies the imbalance between forward and backward probability fluxes of temperature trajectories (Fig. \ref{shaded}(A)). Again, irreversibility of scalar records increases with the lag $\tau$ and here tend to plateu at larger scales ($\tau>1$).

Phase-randomized time series, by comparison, exhibit smaller values of $\langle Z_{\tau} \rangle$ in the inertial range. As one would expect, the excess is thus likely a direct result of the presence of scalar ramps.  The presence of asymmetric patterns in temperature time traces further suggests that in the inertial range scalar turbulence is more time-irreversible than velocity, as confirmed by the larger values of $\langle Z_{\tau} \rangle$ at inertial scales (Fig. \ref{shaded}(B)).  

Time-irreversibility of phase space trajectories was further investigated by testing if a significant difference exists between the probability distribution $p(T'_{\tau} \vert T)$ and $p(-T'_{\tau} \vert T)$. To this end, a Kolmogorov-Smirnov (KS) test was performed at the significance level $0.05$. At every scale $\tau$, results were averaged over different values of $T$ and across runs within the same stability class. The results from the KS test confirm the picture obtained from the relative entropy measure $\langle Z_{\tau} \rangle$: The pdf of forward and backward temperature diverge significantly as the scale $\tau$ increases as shown in figure \ref{shaded}, panels (C) and (D). While this test does not capture the sign of the ramps, the behavior of near neutral runs exhibit some difference from the case of relevant heat flux: near neutral runs appear on average more reversible than unstable runs at the same dimensionless scale $\tau$.

\section*{Discussion and Conclusions}

It is well known that the pdfs of scalar increments develop heavier tails with decreasing scales in the inertial range when compared to their velocity counterparts. The analysis here shows that within the first two decades of the inertial subrange, this buildup of tails also carries the signature of turbulent kinetic energy generation. The direct injection of scalar variance from large scales seem to hinder any universal description of $\Delta T$ statistics within this range of scales. Instead, the pdf of $\Delta T(r)$ for ASL flows appear to be conditional on the value of $\zeta$ at scale $r$. This finding reinforces previous experimental results \cite{lepore2009effect} obtained for a different type of flow (turbulent wake). In this case, the scalar injection mechanism was shown to impact higher order scaling exponents of the temperature structure functions.

This dependence on atmospheric stability regime for $p(\Delta T)$ further suggests that the topology of large eddies, and in particular the presence of ramp-cliff scalar structures, may be responsible for the scale-wise evolution of intermittency and the persistent time directionality at fine scales. This intermittency excess observed in the transition from production to inertial scales is consistent with self-amplification dynamics taking place that further excite the excess of scalar variance injected by the ramps.

However, while measures of intermittency appear to be dependent on the absolute value of $\zeta$, i.e., on the relative magnitude of shear and buoyancy production terms (regardless on the sign of the heat flux), the analysis of asymmetry and time reversibility clearly sense the sign of the heat flux $H$ more than the magnitude of $\zeta$ itself. This effect is arguably a product of the preferential orientation that the external temperature gradient imposes on the scalar ramp-cliffs, as explained by sweep-ejection dynamics. This hypothesis was here further tested by comparisons with synthetic time series that mimic ramp-cliff patterns observed in the scalar time series. The analysis confirmed that much of the observed time irreversibility, as well as its dependence on the sign of $H$, are recovered by these surrogate time series (Fig. \ref{Irrev}).

Our analysis of time directional properties showed that time-irreversible behavior for the scalar is stronger at the large scales of the flow where boundary conditions, and in particular the sign of $H$, determine the orientation and structure of the eddies. At finer scales, time irreversibility as quantified by both $\langle Z_{\tau} \rangle$ and $R_{\tau}$ progressively decreases as advection destroys the preferential eddy orientation imposed by boundary conditions. Note that this behavior is not captured by a simple measure of skewness such as $S^3_T$ (Fig. \ref{Irrev}(A)), which is small at large scales and plateaus in the inertial range consistent with previous experiments \cite{warhaft2000passive} and numerical simulations \cite{celani2000universality}, thus showing that local isotropy is not fully attained at the finer scales examined here.

Turbulent flows exist in a state far from thermodynamic equilibrium, with the flow statistics exhibiting irreversibility. This irreversibility is typically described in terms of fluxes of energy or asymmetries in the pdfs of the fluid velocity increments \cite{falkovich2009symmetries}. Similar methods could be used to describe irreversibility in the scalar field, e.g. using $S^3_T$, and this would imply that the irreversibility of the scalar field is stronger at smaller scales than it is at larger scales. However, in this paper we have used alternative measures to quantify the irreversibility, namely $\langle Z_{\tau} \rangle$ and $R_{\tau}$. These quantities paint a different picture, namely that it is the largest scales, not the smallest (inertial) scales in the scalar field that exhibit the strongest irreversibility. A potential cause for these differing behaviors is that whereas fluxes and quantities such as $S^3_T$ are multi-point, single-time quantities, $\langle Z_{\tau} \rangle$ and $R_{\tau}$ are single-point, multi-time quantities. Thus, these two ways of describing irreversibility provide different perspectives about the nature of irreversibility in turbulence, which involves fields that evolve in both space and time. This difference in perspectives is a topic for future inquiry.

Collectively, the results presented in this paper suggest the following picture for ASL turbulence at the cross-over from production to inertial. Increasing instability in the ASL leads to increases in the mean turbulent kinetic energy dissipation rate (as evidenced by Eq. (\ref{TKE_Budget})) and its spatial autocorrelation function and pdf.  The consequences of this increased dissipation with increased instability has different outcomes for velocity and scalar turbulence.  For velocity, refinements to K41 appear sufficient to explain the observed scaling in the inertial subrange. For scalar turbulence, the picture appears more complicated.  Intermittency buildup with decreasing (inertial) scales is more rapid when compared to their velocity counterparts, and the signature of the temperature variance injection mechanism persists at even the finer scales explored here.  

Turbulence and scalar turbulence are characterized by a constant flux of energy and scalar variance from the scales of production down to dissipation. While early theories hypothesized a cascade only depending on these quantities, experimental evidence to date supports a more complicated picture. The multi-time information encoded in $\langle Z_{\tau} \rangle$ reveal that time-reversibility is not constant across scales, as do the fluxes of information entropy. Probability fluxes forward and backward in time are not balanced in general for air temperature increments, especially at the cross-over from production to inertial. Furthermore, these fluxes carry the signature of the external boundary conditions (i.e. $H$) and show that dissipation rates themselves are not independent of the large-scale dynamics. Although a formal analogy between Eq. (\ref{rel_entr1}) and the thermodynamics of microscopic non-equilibrium steady state systems exists, we stress that in the present application turbulent fluctuations are macroscopic and are the result of non-linear and non-local interactions.

\appendix*
\section{Stable stratification and distortions of the inertial subrange}

In general, stable stratification limits the onset and extent of the inertial subrange given its damping effect in the vertical direction \cite{rorai2015stably}. Here, we show that the scales for which these effects are relevant occur at scales larger than the inertial range examined here. The Ozmidov length scale\cite{ozmidov1965turbulent} (originally suggested by Dougherty \cite{dougherty1961anisotropy} in 1961), is defined as the scale above which buoyancy forces significantly distort the spectrum of turbulence.

\begin{figure}[ht]
\begin{center}
\centerline{\includegraphics[scale=0.8]{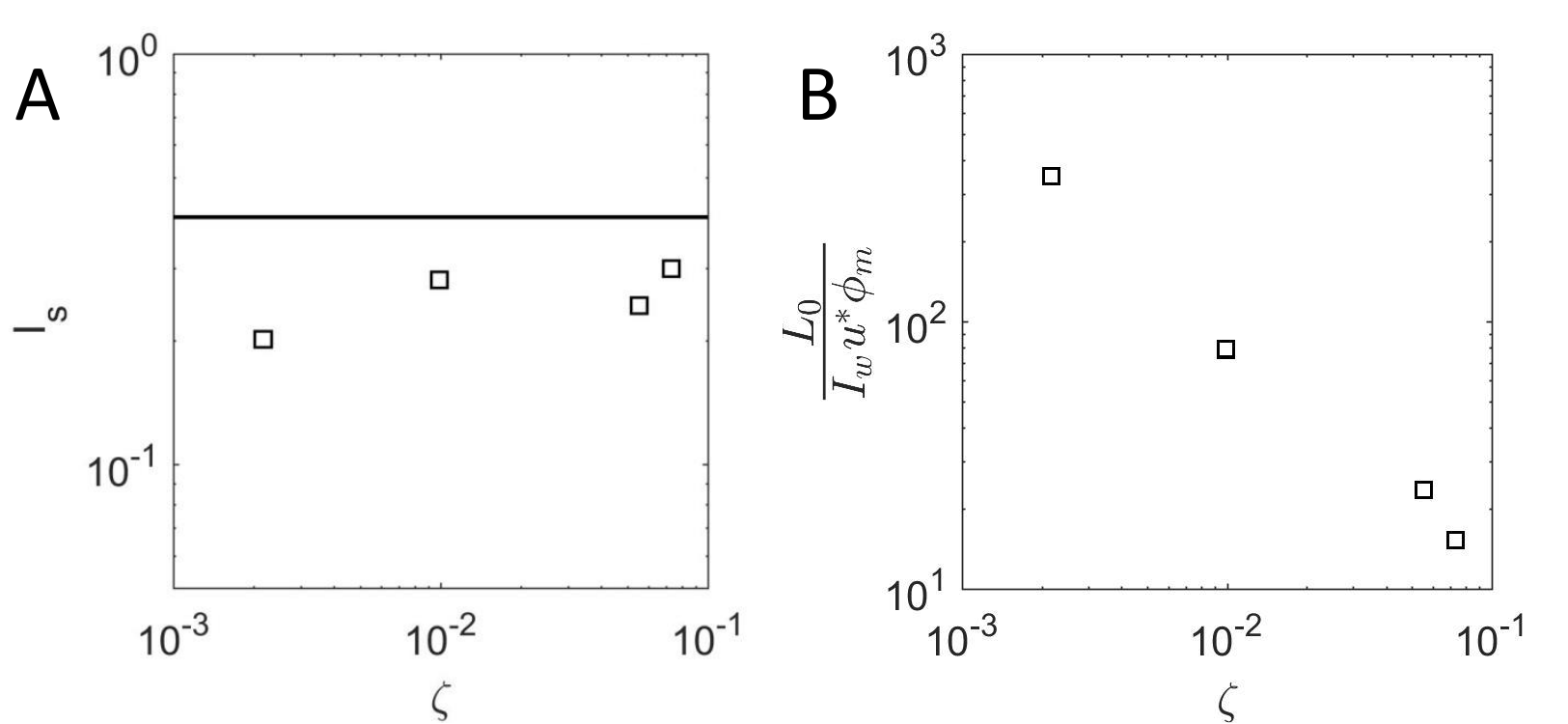}}
\caption{(A) Quantity $Is$ and its expected value 0.4 (black horizontal line) for the 4 stable runs in the dataset. (B) Normalized Ozmidov length for the same runs.}
\label{Ozmidov}
\end{center}
\end{figure}

This length scale, sometimes labeled as the Dougherty-Ozmidov scale, can be expressed as
\begin{equation}
L_0 = \sqrt{\frac{\epsilon}{N^3}},
\end{equation}
where $\epsilon$ is, as before, the mean turbulent kinetic energy  dissipation rate and $N$ is the \emph{Brunt V\"{a}is\"{a}l\"{a}} frequency, defined as
\begin{equation}
N = \sqrt{ \frac{g}{T} \frac{dT}{dz}}.
\end{equation}
In the study used here, no information was provided about the actual mean potential temperature gradient $dT/dz$. However, an approximated estimate of $L_0$ for the runs collected in case of stable atmospheric stratification may be conducted. Note that only 4 runs follow this stability class as runs not meeting strict stationarity requirements were excluded from the analysis (and they were mainly collected in unstable atmospheric conditions). The mean $dT/dz$ was computed using Monin- Obukhov similarity theory as 
\begin{equation}
\frac{dT}{dz} = -\left( \frac{T^*}{K_v z} \right) \phi_T \left( \frac{z}{L} \right)
\end{equation}
where $k_v = 0.41$ is the \emph{von Karman} constant, $z = 5.1$ m is the distance from the ground, $ T^* = \frac{ \langle w'T' \rangle }{u^*}$, and for mildly stable stratification
\begin{equation}
\phi_T = \phi_m = 1 + 4.7 \left( \frac{z}{L} \right).
\end{equation}
The mean turbulent kinetic energy dissipation rate was computed as
\begin{equation}
\epsilon = \frac{   u^{*3} }{k_v z} \left( \phi_m - \frac{z}{L} \right)
\end{equation}
Figure \ref{Ozmidov}(A) shows that the quantity 
\begin{equation}
I_s = \frac{I_w u^* \phi_m}{ k_v z} = constant \simeq 0.4
\end{equation}
is almost constant across runs and exhibits a value slightly lower than the expected $0.4$.

The estimated values of the dimensionless Ozmidov number $ L_0 /  \left( I_w u^* \phi_m \right)$ are reported in Figure \ref{Ozmidov}(B). $L_0$ decreases with increasing stability $\zeta$ as the effect of buoyancy is felt by eddies of sizes progressively smaller. However, the values of the Ozmidov scale are consistently larger than the integral scale of the flow $I_w$ for the 4 stable runs here. Hence, ignoring distortions caused by stable stratification on inertial subrange scales for the aforementioned 4 runs may be deemed plausible.

\section*{Acknowledgments}
E.Z. acknowledges support from the Division of Earth and Ocean Sciences, the Nicholas School of the Environment at Duke University. G.K. acknowledges support from the National Science Foundation (NSF-EAR-1344703, NSF-AGS-1644382, and NSF-DGE-1068871) and from the Department of Energy (DE-SC0011461). Helpful discussions with Marco Marani, Brad Murray and Amilcare Porporato are gratefully acknowledged.

\section*{Disclaimer}
The authors declare no conflict of interest.

  \bibliographystyle{apsrev4-1}
  \bibliography{Turbulence}

\begin{thebibliography}{66}%
\makeatletter
\providecommand \@ifxundefined [1]{%
 \@ifx{#1\undefined}
}%
\providecommand \@ifnum [1]{%
 \ifnum #1\expandafter \@firstoftwo
 \else \expandafter \@secondoftwo
 \fi
}%
\providecommand \@ifx [1]{%
 \ifx #1\expandafter \@firstoftwo
 \else \expandafter \@secondoftwo
 \fi
}%
\providecommand \natexlab [1]{#1}%
\providecommand \enquote  [1]{``#1''}%
\providecommand \bibnamefont  [1]{#1}%
\providecommand \bibfnamefont [1]{#1}%
\providecommand \citenamefont [1]{#1}%
\providecommand \href@noop [0]{\@secondoftwo}%
\providecommand \href [0]{\begingroup \@sanitize@url \@href}%
\providecommand \@href[1]{\@@startlink{#1}\@@href}%
\providecommand \@@href[1]{\endgroup#1\@@endlink}%
\providecommand \@sanitize@url [0]{\catcode `\\12\catcode `\$12\catcode
  `\&12\catcode `\#12\catcode `\^12\catcode `\_12\catcode `\%12\relax}%
\providecommand \@@startlink[1]{}%
\providecommand \@@endlink[0]{}%
\providecommand \url  [0]{\begingroup\@sanitize@url \@url }%
\providecommand \@url [1]{\endgroup\@href {#1}{\urlprefix }}%
\providecommand \urlprefix  [0]{URL }%
\providecommand \Eprint [0]{\href }%
\providecommand \doibase [0]{http://dx.doi.org/}%
\providecommand \selectlanguage [0]{\@gobble}%
\providecommand \bibinfo  [0]{\@secondoftwo}%
\providecommand \bibfield  [0]{\@secondoftwo}%
\providecommand \translation [1]{[#1]}%
\providecommand \BibitemOpen [0]{}%
\providecommand \bibitemStop [0]{}%
\providecommand \bibitemNoStop [0]{.\EOS\space}%
\providecommand \EOS [0]{\spacefactor3000\relax}%
\providecommand \BibitemShut  [1]{\csname bibitem#1\endcsname}%
\let\auto@bib@innerbib\@empty
\bibitem [{\citenamefont {Sreenivasan}(1999)}]{sreenivasan1999fluid}%
  \BibitemOpen
  \bibfield  {author} {\bibinfo {author} {\bibfnamefont {K.~R.}\ \bibnamefont
  {Sreenivasan}},\ }\href@noop {} {\bibfield  {journal} {\bibinfo  {journal}
  {Reviews of Modern Physics}\ }\textbf {\bibinfo {volume} {71}},\ \bibinfo
  {pages} {S383} (\bibinfo {year} {1999})}\BibitemShut {NoStop}%
\bibitem [{\citenamefont {Antonia}\ \emph {et~al.}(1979)\citenamefont
  {Antonia}, \citenamefont {Chambers}, \citenamefont {Friehe},\ and\
  \citenamefont {Van~Atta}}]{antonia1979temperature}%
  \BibitemOpen
  \bibfield  {author} {\bibinfo {author} {\bibfnamefont {R.}~\bibnamefont
  {Antonia}}, \bibinfo {author} {\bibfnamefont {A.}~\bibnamefont {Chambers}},
  \bibinfo {author} {\bibfnamefont {C.}~\bibnamefont {Friehe}}, \ and\ \bibinfo
  {author} {\bibfnamefont {C.}~\bibnamefont {Van~Atta}},\ }\href@noop {}
  {\bibfield  {journal} {\bibinfo  {journal} {Journal of the Atmospheric
  Sciences}\ }\textbf {\bibinfo {volume} {36}},\ \bibinfo {pages} {99}
  (\bibinfo {year} {1979})}\BibitemShut {NoStop}%
\bibitem [{\citenamefont {Shraiman}\ and\ \citenamefont
  {Siggia}(2000)}]{shraiman2000scalar}%
  \BibitemOpen
  \bibfield  {author} {\bibinfo {author} {\bibfnamefont {B.~I.}\ \bibnamefont
  {Shraiman}}\ and\ \bibinfo {author} {\bibfnamefont {E.~D.}\ \bibnamefont
  {Siggia}},\ }\href@noop {} {\bibfield  {journal} {\bibinfo  {journal}
  {Nature}\ }\textbf {\bibinfo {volume} {405}},\ \bibinfo {pages} {639}
  (\bibinfo {year} {2000})}\BibitemShut {NoStop}%
\bibitem [{\citenamefont {Warhaft}(2000)}]{warhaft2000passive}%
  \BibitemOpen
  \bibfield  {author} {\bibinfo {author} {\bibfnamefont {Z.}~\bibnamefont
  {Warhaft}},\ }\href@noop {} {\bibfield  {journal} {\bibinfo  {journal}
  {Annual Review of Fluid Mechanics}\ }\textbf {\bibinfo {volume} {32}},\
  \bibinfo {pages} {203} (\bibinfo {year} {2000})}\BibitemShut {NoStop}%
\bibitem [{\citenamefont {Garratt}(1994)}]{garratt1994review}%
  \BibitemOpen
  \bibfield  {author} {\bibinfo {author} {\bibfnamefont {J.~R.}\ \bibnamefont
  {Garratt}},\ }\href@noop {} {\bibfield  {journal} {\bibinfo  {journal}
  {Earth-Science Reviews}\ }\textbf {\bibinfo {volume} {37}},\ \bibinfo {pages}
  {89} (\bibinfo {year} {1994})}\BibitemShut {NoStop}%
\bibitem [{\citenamefont {Stull}(2012)}]{stull2012introduction}%
  \BibitemOpen
  \bibfield  {author} {\bibinfo {author} {\bibfnamefont {R.~B.}\ \bibnamefont
  {Stull}},\ }\href@noop {} {\emph {\bibinfo {title} {An Introduction to
  Boundary Layer Meteorology}}},\ Vol.~\bibinfo {volume} {13}\ (\bibinfo
  {publisher} {Springer Science \& Business Media},\ \bibinfo {year}
  {2012})\BibitemShut {NoStop}%
\bibitem [{\citenamefont {Kolmogorov}(1941)}]{kolmogorov1941local}%
  \BibitemOpen
  \bibfield  {author} {\bibinfo {author} {\bibfnamefont {A.~N.}\ \bibnamefont
  {Kolmogorov}},\ }in\ \href@noop {} {\emph {\bibinfo {booktitle} {Dokl. Akad.
  Nauk SSSR}}},\ Vol.~\bibinfo {volume} {30}\ (\bibinfo {organization}
  {JSTOR},\ \bibinfo {year} {1941})\ pp.\ \bibinfo {pages}
  {301--305}\BibitemShut {NoStop}%
\bibitem [{\citenamefont {Kraichnan}(1968)}]{kraichnan1968small}%
  \BibitemOpen
  \bibfield  {author} {\bibinfo {author} {\bibfnamefont {R.~H.}\ \bibnamefont
  {Kraichnan}},\ }\href@noop {} {\bibfield  {journal} {\bibinfo  {journal} {The
  Physics of Fluids}\ }\textbf {\bibinfo {volume} {11}},\ \bibinfo {pages}
  {945} (\bibinfo {year} {1968})}\BibitemShut {NoStop}%
\bibitem [{\citenamefont {Kuznetsov}\ \emph {et~al.}(1992)\citenamefont
  {Kuznetsov}, \citenamefont {Praskovsky},\ and\ \citenamefont
  {Sabelnikov}}]{kuznetsov1992fine}%
  \BibitemOpen
  \bibfield  {author} {\bibinfo {author} {\bibfnamefont {V.}~\bibnamefont
  {Kuznetsov}}, \bibinfo {author} {\bibfnamefont {A.}~\bibnamefont
  {Praskovsky}}, \ and\ \bibinfo {author} {\bibfnamefont {V.}~\bibnamefont
  {Sabelnikov}},\ }\href@noop {} {\bibfield  {journal} {\bibinfo  {journal}
  {Journal of Fluid Mechanics}\ }\textbf {\bibinfo {volume} {243}},\ \bibinfo
  {pages} {595} (\bibinfo {year} {1992})}\BibitemShut {NoStop}%
\bibitem [{\citenamefont {Schumacher}\ \emph {et~al.}(2014)\citenamefont
  {Schumacher}, \citenamefont {Scheel}, \citenamefont {Krasnov}, \citenamefont
  {Donzis}, \citenamefont {Yakhot},\ and\ \citenamefont
  {Sreenivasan}}]{schumacher2014small}%
  \BibitemOpen
  \bibfield  {author} {\bibinfo {author} {\bibfnamefont {J.}~\bibnamefont
  {Schumacher}}, \bibinfo {author} {\bibfnamefont {J.~D.}\ \bibnamefont
  {Scheel}}, \bibinfo {author} {\bibfnamefont {D.}~\bibnamefont {Krasnov}},
  \bibinfo {author} {\bibfnamefont {D.~A.}\ \bibnamefont {Donzis}}, \bibinfo
  {author} {\bibfnamefont {V.}~\bibnamefont {Yakhot}}, \ and\ \bibinfo {author}
  {\bibfnamefont {K.~R.}\ \bibnamefont {Sreenivasan}},\ }\href@noop {}
  {\bibfield  {journal} {\bibinfo  {journal} {Proceedings of the National
  Academy of Sciences}\ }\textbf {\bibinfo {volume} {111}},\ \bibinfo {pages}
  {10961} (\bibinfo {year} {2014})}\BibitemShut {NoStop}%
\bibitem [{\citenamefont {Yeung}\ \emph {et~al.}(2015)\citenamefont {Yeung},
  \citenamefont {Zhai},\ and\ \citenamefont {Sreenivasan}}]{yeung2015extreme}%
  \BibitemOpen
  \bibfield  {author} {\bibinfo {author} {\bibfnamefont {P.}~\bibnamefont
  {Yeung}}, \bibinfo {author} {\bibfnamefont {X.}~\bibnamefont {Zhai}}, \ and\
  \bibinfo {author} {\bibfnamefont {K.~R.}\ \bibnamefont {Sreenivasan}},\
  }\href@noop {} {\bibfield  {journal} {\bibinfo  {journal} {Proceedings of the
  National Academy of Sciences}\ }\textbf {\bibinfo {volume} {112}},\ \bibinfo
  {pages} {12633} (\bibinfo {year} {2015})}\BibitemShut {NoStop}%
\bibitem [{\citenamefont {Katul}\ \emph {et~al.}(2015)\citenamefont {Katul},
  \citenamefont {Manes}, \citenamefont {Porporato}, \citenamefont {Bou-Zeid},\
  and\ \citenamefont {Chamecki}}]{katul2015bottlenecks}%
  \BibitemOpen
  \bibfield  {author} {\bibinfo {author} {\bibfnamefont {G.~G.}\ \bibnamefont
  {Katul}}, \bibinfo {author} {\bibfnamefont {C.}~\bibnamefont {Manes}},
  \bibinfo {author} {\bibfnamefont {A.}~\bibnamefont {Porporato}}, \bibinfo
  {author} {\bibfnamefont {E.}~\bibnamefont {Bou-Zeid}}, \ and\ \bibinfo
  {author} {\bibfnamefont {M.}~\bibnamefont {Chamecki}},\ }\href@noop {}
  {\bibfield  {journal} {\bibinfo  {journal} {Physical Review E}\ }\textbf
  {\bibinfo {volume} {92}},\ \bibinfo {pages} {033009} (\bibinfo {year}
  {2015})}\BibitemShut {NoStop}%
\bibitem [{\citenamefont {Katul}\ \emph {et~al.}(2006)\citenamefont {Katul},
  \citenamefont {Porporato}, \citenamefont {Cava},\ and\ \citenamefont
  {Siqueira}}]{katul2006analysis}%
  \BibitemOpen
  \bibfield  {author} {\bibinfo {author} {\bibfnamefont {G.}~\bibnamefont
  {Katul}}, \bibinfo {author} {\bibfnamefont {A.}~\bibnamefont {Porporato}},
  \bibinfo {author} {\bibfnamefont {D.}~\bibnamefont {Cava}}, \ and\ \bibinfo
  {author} {\bibfnamefont {M.}~\bibnamefont {Siqueira}},\ }\href@noop {}
  {\bibfield  {journal} {\bibinfo  {journal} {Physica D: Nonlinear Phenomena}\
  }\textbf {\bibinfo {volume} {215}},\ \bibinfo {pages} {117} (\bibinfo {year}
  {2006})}\BibitemShut {NoStop}%
\bibitem [{\citenamefont {Reynolds}(2012)}]{reynolds2012gusts}%
  \BibitemOpen
  \bibfield  {author} {\bibinfo {author} {\bibfnamefont {A.}~\bibnamefont
  {Reynolds}},\ }\href@noop {} {\bibfield  {journal} {\bibinfo  {journal}
  {Physica A: Statistical Mechanics and its Applications}\ }\textbf {\bibinfo
  {volume} {391}},\ \bibinfo {pages} {5059} (\bibinfo {year}
  {2012})}\BibitemShut {NoStop}%
\bibitem [{\citenamefont {Xu}\ \emph {et~al.}(2014)\citenamefont {Xu},
  \citenamefont {Pumir}, \citenamefont {Falkovich}, \citenamefont
  {Bodenschatz}, \citenamefont {Shats}, \citenamefont {Xia}, \citenamefont
  {Francois},\ and\ \citenamefont {Boffetta}}]{xu2014flight}%
  \BibitemOpen
  \bibfield  {author} {\bibinfo {author} {\bibfnamefont {H.}~\bibnamefont
  {Xu}}, \bibinfo {author} {\bibfnamefont {A.}~\bibnamefont {Pumir}}, \bibinfo
  {author} {\bibfnamefont {G.}~\bibnamefont {Falkovich}}, \bibinfo {author}
  {\bibfnamefont {E.}~\bibnamefont {Bodenschatz}}, \bibinfo {author}
  {\bibfnamefont {M.}~\bibnamefont {Shats}}, \bibinfo {author} {\bibfnamefont
  {H.}~\bibnamefont {Xia}}, \bibinfo {author} {\bibfnamefont {N.}~\bibnamefont
  {Francois}}, \ and\ \bibinfo {author} {\bibfnamefont {G.}~\bibnamefont
  {Boffetta}},\ }\href@noop {} {\bibfield  {journal} {\bibinfo  {journal}
  {Proceedings of the National Academy of Sciences}\ }\textbf {\bibinfo
  {volume} {111}},\ \bibinfo {pages} {7558} (\bibinfo {year}
  {2014})}\BibitemShut {NoStop}%
\bibitem [{\citenamefont {Falkovich}\ \emph {et~al.}(2001)\citenamefont
  {Falkovich}, \citenamefont {Gawȩdzki},\ and\ \citenamefont
  {Vergassola}}]{falkovich2001particles}%
  \BibitemOpen
  \bibfield  {author} {\bibinfo {author} {\bibfnamefont {G.}~\bibnamefont
  {Falkovich}}, \bibinfo {author} {\bibfnamefont {K.}~\bibnamefont
  {Gawȩdzki}}, \ and\ \bibinfo {author} {\bibfnamefont {M.}~\bibnamefont
  {Vergassola}},\ }\href@noop {} {\bibfield  {journal} {\bibinfo  {journal}
  {Reviews of Modern Physics}\ }\textbf {\bibinfo {volume} {73}},\ \bibinfo
  {pages} {913} (\bibinfo {year} {2001})}\BibitemShut {NoStop}%
\bibitem [{\citenamefont {Meneveau}\ \emph {et~al.}(1996)\citenamefont
  {Meneveau}, \citenamefont {Lund},\ and\ \citenamefont
  {Cabot}}]{meneveau1996lagrangian}%
  \BibitemOpen
  \bibfield  {author} {\bibinfo {author} {\bibfnamefont {C.}~\bibnamefont
  {Meneveau}}, \bibinfo {author} {\bibfnamefont {T.~S.}\ \bibnamefont {Lund}},
  \ and\ \bibinfo {author} {\bibfnamefont {W.~H.}\ \bibnamefont {Cabot}},\
  }\href@noop {} {\bibfield  {journal} {\bibinfo  {journal} {Journal of Fluid
  Mechanics}\ }\textbf {\bibinfo {volume} {319}},\ \bibinfo {pages} {353}
  (\bibinfo {year} {1996})}\BibitemShut {NoStop}%
\bibitem [{\citenamefont {Port{\'e}-Agel}\ \emph {et~al.}(2000)\citenamefont
  {Port{\'e}-Agel}, \citenamefont {Meneveau},\ and\ \citenamefont
  {Parlange}}]{porte2000scale}%
  \BibitemOpen
  \bibfield  {author} {\bibinfo {author} {\bibfnamefont {F.}~\bibnamefont
  {Port{\'e}-Agel}}, \bibinfo {author} {\bibfnamefont {C.}~\bibnamefont
  {Meneveau}}, \ and\ \bibinfo {author} {\bibfnamefont {M.~B.}\ \bibnamefont
  {Parlange}},\ }\href@noop {} {\bibfield  {journal} {\bibinfo  {journal}
  {Journal of Fluid Mechanics}\ }\textbf {\bibinfo {volume} {415}},\ \bibinfo
  {pages} {261} (\bibinfo {year} {2000})}\BibitemShut {NoStop}%
\bibitem [{\citenamefont {Higgins}\ \emph {et~al.}(2003)\citenamefont
  {Higgins}, \citenamefont {Parlange},\ and\ \citenamefont
  {Meneveau}}]{higgins2003alignment}%
  \BibitemOpen
  \bibfield  {author} {\bibinfo {author} {\bibfnamefont {C.~W.}\ \bibnamefont
  {Higgins}}, \bibinfo {author} {\bibfnamefont {M.~B.}\ \bibnamefont
  {Parlange}}, \ and\ \bibinfo {author} {\bibfnamefont {C.}~\bibnamefont
  {Meneveau}},\ }\href@noop {} {\bibfield  {journal} {\bibinfo  {journal}
  {Boundary-Layer Meteorology}\ }\textbf {\bibinfo {volume} {109}},\ \bibinfo
  {pages} {59} (\bibinfo {year} {2003})}\BibitemShut {NoStop}%
\bibitem [{\citenamefont {Stoll}\ and\ \citenamefont
  {Port{\'e}-Agel}(2006)}]{stoll2006dynamic}%
  \BibitemOpen
  \bibfield  {author} {\bibinfo {author} {\bibfnamefont {R.}~\bibnamefont
  {Stoll}}\ and\ \bibinfo {author} {\bibfnamefont {F.}~\bibnamefont
  {Port{\'e}-Agel}},\ }\href@noop {} {\bibfield  {journal} {\bibinfo  {journal}
  {Water Resources Research}\ }\textbf {\bibinfo {volume} {42}} (\bibinfo
  {year} {2006})}\BibitemShut {NoStop}%
\bibitem [{\citenamefont {Katul}\ \emph {et~al.}(2011)\citenamefont {Katul},
  \citenamefont {Konings},\ and\ \citenamefont {Porporato}}]{katul2011mean}%
  \BibitemOpen
  \bibfield  {author} {\bibinfo {author} {\bibfnamefont {G.~G.}\ \bibnamefont
  {Katul}}, \bibinfo {author} {\bibfnamefont {A.~G.}\ \bibnamefont {Konings}},
  \ and\ \bibinfo {author} {\bibfnamefont {A.}~\bibnamefont {Porporato}},\
  }\href@noop {} {\bibfield  {journal} {\bibinfo  {journal} {Physical Review
  Letters}\ }\textbf {\bibinfo {volume} {107}},\ \bibinfo {pages} {268502}
  (\bibinfo {year} {2011})}\BibitemShut {NoStop}%
\bibitem [{\citenamefont {Katul}\ \emph {et~al.}(2013)\citenamefont {Katul},
  \citenamefont {Li}, \citenamefont {Chamecki},\ and\ \citenamefont
  {Bou-Zeid}}]{katul2013mean}%
  \BibitemOpen
  \bibfield  {author} {\bibinfo {author} {\bibfnamefont {G.~G.}\ \bibnamefont
  {Katul}}, \bibinfo {author} {\bibfnamefont {D.}~\bibnamefont {Li}}, \bibinfo
  {author} {\bibfnamefont {M.}~\bibnamefont {Chamecki}}, \ and\ \bibinfo
  {author} {\bibfnamefont {E.}~\bibnamefont {Bou-Zeid}},\ }\href@noop {}
  {\bibfield  {journal} {\bibinfo  {journal} {Physical Review E}\ }\textbf
  {\bibinfo {volume} {87}},\ \bibinfo {pages} {023004} (\bibinfo {year}
  {2013})}\BibitemShut {NoStop}%
\bibitem [{\citenamefont {Li}\ \emph {et~al.}(2012)\citenamefont {Li},
  \citenamefont {Katul},\ and\ \citenamefont {Bou-Zeid}}]{li2012mean}%
  \BibitemOpen
  \bibfield  {author} {\bibinfo {author} {\bibfnamefont {D.}~\bibnamefont
  {Li}}, \bibinfo {author} {\bibfnamefont {G.~G.}\ \bibnamefont {Katul}}, \
  and\ \bibinfo {author} {\bibfnamefont {E.}~\bibnamefont {Bou-Zeid}},\
  }\href@noop {} {\bibfield  {journal} {\bibinfo  {journal} {Physics of
  Fluids}\ }\textbf {\bibinfo {volume} {24}},\ \bibinfo {pages} {105105}
  (\bibinfo {year} {2012})}\BibitemShut {NoStop}%
\bibitem [{\citenamefont {Katul}\ \emph {et~al.}(2014)\citenamefont {Katul},
  \citenamefont {Porporato}, \citenamefont {Shah},\ and\ \citenamefont
  {Bou-Zeid}}]{katul2014two}%
  \BibitemOpen
  \bibfield  {author} {\bibinfo {author} {\bibfnamefont {G.~G.}\ \bibnamefont
  {Katul}}, \bibinfo {author} {\bibfnamefont {A.}~\bibnamefont {Porporato}},
  \bibinfo {author} {\bibfnamefont {S.}~\bibnamefont {Shah}}, \ and\ \bibinfo
  {author} {\bibfnamefont {E.}~\bibnamefont {Bou-Zeid}},\ }\href@noop {}
  {\bibfield  {journal} {\bibinfo  {journal} {Physical Review E}\ }\textbf
  {\bibinfo {volume} {89}},\ \bibinfo {pages} {023007} (\bibinfo {year}
  {2014})}\BibitemShut {NoStop}%
\bibitem [{\citenamefont {Li}\ \emph {et~al.}(2015)\citenamefont {Li},
  \citenamefont {Katul},\ and\ \citenamefont
  {Zilitinkevich}}]{li2015revisiting}%
  \BibitemOpen
  \bibfield  {author} {\bibinfo {author} {\bibfnamefont {D.}~\bibnamefont
  {Li}}, \bibinfo {author} {\bibfnamefont {G.~G.}\ \bibnamefont {Katul}}, \
  and\ \bibinfo {author} {\bibfnamefont {S.~S.}\ \bibnamefont
  {Zilitinkevich}},\ }\href@noop {} {\bibfield  {journal} {\bibinfo  {journal}
  {Journal of the Atmospheric Sciences}\ }\textbf {\bibinfo {volume} {72}},\
  \bibinfo {pages} {2394} (\bibinfo {year} {2015})}\BibitemShut {NoStop}%
\bibitem [{\citenamefont {Monin}\ and\ \citenamefont
  {Obukhov}(1954)}]{monin1954basic}%
  \BibitemOpen
  \bibfield  {author} {\bibinfo {author} {\bibfnamefont {A.}~\bibnamefont
  {Monin}}\ and\ \bibinfo {author} {\bibfnamefont {A.}~\bibnamefont
  {Obukhov}},\ }\href@noop {} {\bibfield  {journal} {\bibinfo  {journal}
  {Contrib. Geophys. Inst. Acad. Sci. USSR}\ }\textbf {\bibinfo {volume}
  {151}},\ \bibinfo {pages} {e187} (\bibinfo {year} {1954})}\BibitemShut
  {NoStop}%
\bibitem [{\citenamefont {Chen}\ \emph {et~al.}(1997)\citenamefont {Chen},
  \citenamefont {Novak}, \citenamefont {Black},\ and\ \citenamefont
  {Lee}}]{chen1997coherent}%
  \BibitemOpen
  \bibfield  {author} {\bibinfo {author} {\bibfnamefont {W.}~\bibnamefont
  {Chen}}, \bibinfo {author} {\bibfnamefont {M.~D.}\ \bibnamefont {Novak}},
  \bibinfo {author} {\bibfnamefont {T.~A.}\ \bibnamefont {Black}}, \ and\
  \bibinfo {author} {\bibfnamefont {X.}~\bibnamefont {Lee}},\ }\href@noop {}
  {\bibfield  {journal} {\bibinfo  {journal} {Boundary-Layer Meteorology}\
  }\textbf {\bibinfo {volume} {84}},\ \bibinfo {pages} {99} (\bibinfo {year}
  {1997})}\BibitemShut {NoStop}%
\bibitem [{\citenamefont {Thomas}\ and\ \citenamefont
  {Foken}(2007)}]{thomas2007organised}%
  \BibitemOpen
  \bibfield  {author} {\bibinfo {author} {\bibfnamefont {C.}~\bibnamefont
  {Thomas}}\ and\ \bibinfo {author} {\bibfnamefont {T.}~\bibnamefont {Foken}},\
  }\href@noop {} {\bibfield  {journal} {\bibinfo  {journal} {Boundary-Layer
  Meteorology}\ }\textbf {\bibinfo {volume} {122}},\ \bibinfo {pages} {123}
  (\bibinfo {year} {2007})}\BibitemShut {NoStop}%
\bibitem [{\citenamefont {Katul}\ \emph {et~al.}(1997)\citenamefont {Katul},
  \citenamefont {Hsieh},\ and\ \citenamefont {Sigmon}}]{katul1997energy}%
  \BibitemOpen
  \bibfield  {author} {\bibinfo {author} {\bibfnamefont {G.}~\bibnamefont
  {Katul}}, \bibinfo {author} {\bibfnamefont {C.-I.}\ \bibnamefont {Hsieh}}, \
  and\ \bibinfo {author} {\bibfnamefont {J.}~\bibnamefont {Sigmon}},\
  }\href@noop {} {\bibfield  {journal} {\bibinfo  {journal} {Boundary-Layer
  Meteorology}\ }\textbf {\bibinfo {volume} {82}},\ \bibinfo {pages} {49}
  (\bibinfo {year} {1997})}\BibitemShut {NoStop}%
\bibitem [{\citenamefont {Obukhov}(1949)}]{obukhov1949local}%
  \BibitemOpen
  \bibfield  {author} {\bibinfo {author} {\bibfnamefont {A.}~\bibnamefont
  {Obukhov}},\ }in\ \href@noop {} {\emph {\bibinfo {booktitle} {Dokl. Akad.
  Nauk. SSSR}}},\ Vol.~\bibinfo {volume} {67}\ (\bibinfo {year} {1949})\ pp.\
  \bibinfo {pages} {643--646}\BibitemShut {NoStop}%
\bibitem [{\citenamefont {Sreenivasan}(1991)}]{sreenivasan1991local}%
  \BibitemOpen
  \bibfield  {author} {\bibinfo {author} {\bibfnamefont {K.}~\bibnamefont
  {Sreenivasan}},\ }in\ \href@noop {} {\emph {\bibinfo {booktitle} {Proceedings
  of the Royal Society of London A: Mathematical, Physical and Engineering
  Sciences}}},\ Vol.\ \bibinfo {volume} {434}\ (\bibinfo {organization} {The
  Royal Society},\ \bibinfo {year} {1991})\ pp.\ \bibinfo {pages}
  {165--182}\BibitemShut {NoStop}%
\bibitem [{\citenamefont {Kolmogorov}(1962)}]{kolmogorov1962refinement}%
  \BibitemOpen
  \bibfield  {author} {\bibinfo {author} {\bibfnamefont {A.~N.}\ \bibnamefont
  {Kolmogorov}},\ }\href@noop {} {\bibfield  {journal} {\bibinfo  {journal}
  {Journal of Fluid Mechanics}\ }\textbf {\bibinfo {volume} {13}},\ \bibinfo
  {pages} {82} (\bibinfo {year} {1962})}\BibitemShut {NoStop}%
\bibitem [{\citenamefont {Katul}\ \emph {et~al.}(1994)\citenamefont {Katul},
  \citenamefont {Parlange},\ and\ \citenamefont
  {Chu}}]{katul1994intermittency}%
  \BibitemOpen
  \bibfield  {author} {\bibinfo {author} {\bibfnamefont {G.~G.}\ \bibnamefont
  {Katul}}, \bibinfo {author} {\bibfnamefont {M.~B.}\ \bibnamefont {Parlange}},
  \ and\ \bibinfo {author} {\bibfnamefont {C.~R.}\ \bibnamefont {Chu}},\
  }\href@noop {} {\bibfield  {journal} {\bibinfo  {journal} {Physics of
  Fluids}\ }\textbf {\bibinfo {volume} {6}},\ \bibinfo {pages} {2480} (\bibinfo
  {year} {1994})}\BibitemShut {NoStop}%
\bibitem [{\citenamefont {Katul}\ \emph {et~al.}(2001)\citenamefont {Katul},
  \citenamefont {Vidakovic},\ and\ \citenamefont
  {Albertson}}]{katul2001estimating}%
  \BibitemOpen
  \bibfield  {author} {\bibinfo {author} {\bibfnamefont {G.}~\bibnamefont
  {Katul}}, \bibinfo {author} {\bibfnamefont {B.}~\bibnamefont {Vidakovic}}, \
  and\ \bibinfo {author} {\bibfnamefont {J.}~\bibnamefont {Albertson}},\
  }\href@noop {} {\bibfield  {journal} {\bibinfo  {journal} {Physics of
  Fluids}\ }\textbf {\bibinfo {volume} {13}},\ \bibinfo {pages} {241} (\bibinfo
  {year} {2001})}\BibitemShut {NoStop}%
\bibitem [{\citenamefont {Pope}\ and\ \citenamefont
  {Ching}(1993)}]{pope1993stationary}%
  \BibitemOpen
  \bibfield  {author} {\bibinfo {author} {\bibfnamefont {S.}~\bibnamefont
  {Pope}}\ and\ \bibinfo {author} {\bibfnamefont {E.~S.}\ \bibnamefont
  {Ching}},\ }\href@noop {} {\bibfield  {journal} {\bibinfo  {journal} {Physics
  of Fluids A: Fluid Dynamics}\ }\textbf {\bibinfo {volume} {5}},\ \bibinfo
  {pages} {1529} (\bibinfo {year} {1993})}\BibitemShut {NoStop}%
\bibitem [{\citenamefont {Sinai}\ and\ \citenamefont
  {Yakhot}(1989)}]{sinai1989limiting}%
  \BibitemOpen
  \bibfield  {author} {\bibinfo {author} {\bibfnamefont {Y.~G.}\ \bibnamefont
  {Sinai}}\ and\ \bibinfo {author} {\bibfnamefont {V.}~\bibnamefont {Yakhot}},\
  }\href@noop {} {\bibfield  {journal} {\bibinfo  {journal} {Physical Review
  Letters}\ }\textbf {\bibinfo {volume} {63}},\ \bibinfo {pages} {1962}
  (\bibinfo {year} {1989})}\BibitemShut {NoStop}%
\bibitem [{\citenamefont {Ching}(1993)}]{ching1993probability}%
  \BibitemOpen
  \bibfield  {author} {\bibinfo {author} {\bibfnamefont {E.~S.}\ \bibnamefont
  {Ching}},\ }\href@noop {} {\bibfield  {journal} {\bibinfo  {journal}
  {Physical Review Letters}\ }\textbf {\bibinfo {volume} {70}},\ \bibinfo
  {pages} {283} (\bibinfo {year} {1993})}\BibitemShut {NoStop}%
\bibitem [{\citenamefont {Gardiner}(p124)}]{gardiner1985stochastic}%
  \BibitemOpen
  \bibfield  {author} {\bibinfo {author} {\bibfnamefont {C.~W.}\ \bibnamefont
  {Gardiner}},\ }\href@noop {} {\emph {\bibinfo {title} {Handbook of Stochastic
  Methods}}}\ (\bibinfo  {publisher} {Springer-Verlag, Berlin--Heidelberg--New
  York--Tokyo},\ \bibinfo {year} {1985, p.124})\BibitemShut {NoStop}%
\bibitem [{\citenamefont {Porporato}\ \emph {et~al.}(2011)\citenamefont
  {Porporato}, \citenamefont {Kramer}, \citenamefont {Cassiani}, \citenamefont
  {Daly},\ and\ \citenamefont {Mattingly}}]{porporato2011local}%
  \BibitemOpen
  \bibfield  {author} {\bibinfo {author} {\bibfnamefont {A.}~\bibnamefont
  {Porporato}}, \bibinfo {author} {\bibfnamefont {P.}~\bibnamefont {Kramer}},
  \bibinfo {author} {\bibfnamefont {M.}~\bibnamefont {Cassiani}}, \bibinfo
  {author} {\bibfnamefont {E.}~\bibnamefont {Daly}}, \ and\ \bibinfo {author}
  {\bibfnamefont {J.}~\bibnamefont {Mattingly}},\ }\href@noop {} {\bibfield
  {journal} {\bibinfo  {journal} {Physical Review E}\ }\textbf {\bibinfo
  {volume} {84}},\ \bibinfo {pages} {041142} (\bibinfo {year}
  {2011})}\BibitemShut {NoStop}%
\bibitem [{\citenamefont {Frisch}\ and\ \citenamefont
  {Sornette}(1997)}]{frisch1997extreme}%
  \BibitemOpen
  \bibfield  {author} {\bibinfo {author} {\bibfnamefont {U.}~\bibnamefont
  {Frisch}}\ and\ \bibinfo {author} {\bibfnamefont {D.}~\bibnamefont
  {Sornette}},\ }\href@noop {} {\bibfield  {journal} {\bibinfo  {journal}
  {Journal de Physique I}\ }\textbf {\bibinfo {volume} {7}},\ \bibinfo {pages}
  {1155} (\bibinfo {year} {1997})}\BibitemShut {NoStop}%
\bibitem [{\citenamefont {Tsallis}(1988)}]{tsallis1988possible}%
  \BibitemOpen
  \bibfield  {author} {\bibinfo {author} {\bibfnamefont {C.}~\bibnamefont
  {Tsallis}},\ }\href@noop {} {\bibfield  {journal} {\bibinfo  {journal}
  {Journal of Statistical Physics}\ }\textbf {\bibinfo {volume} {52}},\
  \bibinfo {pages} {479} (\bibinfo {year} {1988})}\BibitemShut {NoStop}%
\bibitem [{\citenamefont {Tsallis}\ \emph {et~al.}(1995)\citenamefont
  {Tsallis}, \citenamefont {Levy}, \citenamefont {Souza},\ and\ \citenamefont
  {Maynard}}]{tsallis1995statistical}%
  \BibitemOpen
  \bibfield  {author} {\bibinfo {author} {\bibfnamefont {C.}~\bibnamefont
  {Tsallis}}, \bibinfo {author} {\bibfnamefont {S.~V.}\ \bibnamefont {Levy}},
  \bibinfo {author} {\bibfnamefont {A.~M.}\ \bibnamefont {Souza}}, \ and\
  \bibinfo {author} {\bibfnamefont {R.}~\bibnamefont {Maynard}},\ }\href@noop
  {} {\bibfield  {journal} {\bibinfo  {journal} {Physical Review Letters}\
  }\textbf {\bibinfo {volume} {75}},\ \bibinfo {pages} {3589} (\bibinfo {year}
  {1995})}\BibitemShut {NoStop}%
\bibitem [{\citenamefont {Shi}\ \emph {et~al.}(2005)\citenamefont {Shi},
  \citenamefont {Vidakovic}, \citenamefont {Katul},\ and\ \citenamefont
  {Albertson}}]{shi2005assessing}%
  \BibitemOpen
  \bibfield  {author} {\bibinfo {author} {\bibfnamefont {B.}~\bibnamefont
  {Shi}}, \bibinfo {author} {\bibfnamefont {B.}~\bibnamefont {Vidakovic}},
  \bibinfo {author} {\bibfnamefont {G.~G.}\ \bibnamefont {Katul}}, \ and\
  \bibinfo {author} {\bibfnamefont {J.~D.}\ \bibnamefont {Albertson}},\
  }\href@noop {} {\bibfield  {journal} {\bibinfo  {journal} {Physics of
  Fluids}\ }\textbf {\bibinfo {volume} {17}},\ \bibinfo {pages} {055104}
  (\bibinfo {year} {2005})}\BibitemShut {NoStop}%
\bibitem [{\citenamefont {Gotoh}\ and\ \citenamefont
  {Kraichnan}(2004)}]{gotoh2004turbulence}%
  \BibitemOpen
  \bibfield  {author} {\bibinfo {author} {\bibfnamefont {T.}~\bibnamefont
  {Gotoh}}\ and\ \bibinfo {author} {\bibfnamefont {R.~H.}\ \bibnamefont
  {Kraichnan}},\ }\href@noop {} {\bibfield  {journal} {\bibinfo  {journal}
  {Physica D: Nonlinear Phenomena}\ }\textbf {\bibinfo {volume} {193}},\
  \bibinfo {pages} {231} (\bibinfo {year} {2004})}\BibitemShut {NoStop}%
\bibitem [{\citenamefont {Ramos}\ \emph {et~al.}(2001)\citenamefont {Ramos},
  \citenamefont {Rosa}, \citenamefont {Neto}, \citenamefont {Bolzan},
  \citenamefont {S{\'a}},\ and\ \citenamefont {Velho}}]{Ramos2001non}%
  \BibitemOpen
  \bibfield  {author} {\bibinfo {author} {\bibfnamefont {F.~M.}\ \bibnamefont
  {Ramos}}, \bibinfo {author} {\bibfnamefont {R.~R.}\ \bibnamefont {Rosa}},
  \bibinfo {author} {\bibfnamefont {C.~R.}\ \bibnamefont {Neto}}, \bibinfo
  {author} {\bibfnamefont {M.~J.}\ \bibnamefont {Bolzan}}, \bibinfo {author}
  {\bibfnamefont {L.~D.~A.}\ \bibnamefont {S{\'a}}}, \ and\ \bibinfo {author}
  {\bibfnamefont {H.~F.~C.}\ \bibnamefont {Velho}},\ }\href@noop {} {\bibfield
  {journal} {\bibinfo  {journal} {Physica A: Statistical Mechanics and its
  Applications}\ }\textbf {\bibinfo {volume} {295}},\ \bibinfo {pages} {250}
  (\bibinfo {year} {2001})}\BibitemShut {NoStop}%
\bibitem [{\citenamefont {Arimitsu}\ and\ \citenamefont
  {Arimitsu}(2002)}]{arimitsu2002tsallis}%
  \BibitemOpen
  \bibfield  {author} {\bibinfo {author} {\bibfnamefont {T.}~\bibnamefont
  {Arimitsu}}\ and\ \bibinfo {author} {\bibfnamefont {N.}~\bibnamefont
  {Arimitsu}},\ }\href@noop {} {\bibfield  {journal} {\bibinfo  {journal}
  {Chaos, Solitons \& Fractals}\ }\textbf {\bibinfo {volume} {13}},\ \bibinfo
  {pages} {479} (\bibinfo {year} {2002})}\BibitemShut {NoStop}%
\bibitem [{\citenamefont {Bolzan}\ \emph {et~al.}(2002)\citenamefont {Bolzan},
  \citenamefont {Ramos}, \citenamefont {S{\'a}}, \citenamefont
  {Rodrigues~Neto},\ and\ \citenamefont {Rosa}}]{bolzan2002analysis}%
  \BibitemOpen
  \bibfield  {author} {\bibinfo {author} {\bibfnamefont {M.~J.}\ \bibnamefont
  {Bolzan}}, \bibinfo {author} {\bibfnamefont {F.~M.}\ \bibnamefont {Ramos}},
  \bibinfo {author} {\bibfnamefont {L.~D.}\ \bibnamefont {S{\'a}}}, \bibinfo
  {author} {\bibfnamefont {C.}~\bibnamefont {Rodrigues~Neto}}, \ and\ \bibinfo
  {author} {\bibfnamefont {R.~R.}\ \bibnamefont {Rosa}},\ }\href@noop {}
  {\bibfield  {journal} {\bibinfo  {journal} {Journal of Geophysical Research:
  Atmospheres}\ }\textbf {\bibinfo {volume} {107}} (\bibinfo {year}
  {2002})}\BibitemShut {NoStop}%
\bibitem [{\citenamefont {Mydlarski}\ \emph {et~al.}(1998)\citenamefont
  {Mydlarski}, \citenamefont {Pumir}, \citenamefont {Shraiman}, \citenamefont
  {Siggia},\ and\ \citenamefont {Warhaft}}]{mydlarski1998structures}%
  \BibitemOpen
  \bibfield  {author} {\bibinfo {author} {\bibfnamefont {L.}~\bibnamefont
  {Mydlarski}}, \bibinfo {author} {\bibfnamefont {A.}~\bibnamefont {Pumir}},
  \bibinfo {author} {\bibfnamefont {B.~I.}\ \bibnamefont {Shraiman}}, \bibinfo
  {author} {\bibfnamefont {E.~D.}\ \bibnamefont {Siggia}}, \ and\ \bibinfo
  {author} {\bibfnamefont {Z.}~\bibnamefont {Warhaft}},\ }\href@noop {}
  {\bibfield  {journal} {\bibinfo  {journal} {Physical review letters}\
  }\textbf {\bibinfo {volume} {81}},\ \bibinfo {pages} {4373} (\bibinfo {year}
  {1998})}\BibitemShut {NoStop}%
\bibitem [{\citenamefont {Lawrance}(1991)}]{lawrance1991directionality}%
  \BibitemOpen
  \bibfield  {author} {\bibinfo {author} {\bibfnamefont {A.}~\bibnamefont
  {Lawrance}},\ }\href@noop {} {\bibfield  {journal} {\bibinfo  {journal}
  {International Statistical Review/Revue Internationale de Statistique}\ ,\
  \bibinfo {pages} {67}} (\bibinfo {year} {1991})}\BibitemShut {NoStop}%
\bibitem [{\citenamefont {Cover}\ and\ \citenamefont {Thomas}(
  p18)}]{cover2012elements}%
  \BibitemOpen
  \bibfield  {author} {\bibinfo {author} {\bibfnamefont {T.~M.}\ \bibnamefont
  {Cover}}\ and\ \bibinfo {author} {\bibfnamefont {J.~A.}\ \bibnamefont
  {Thomas}},\ }\href@noop {} {\emph {\bibinfo {title} {Elements of Information
  Theory}}}\ (\bibinfo  {publisher} {John Wiley \& Sons},\ \bibinfo {year}
  {2012, p.18})\BibitemShut {NoStop}%
\bibitem [{\citenamefont {Porporato}\ \emph {et~al.}(2007)\citenamefont
  {Porporato}, \citenamefont {Rigby},\ and\ \citenamefont
  {Daly}}]{porporato2007irreversibility}%
  \BibitemOpen
  \bibfield  {author} {\bibinfo {author} {\bibfnamefont {A.}~\bibnamefont
  {Porporato}}, \bibinfo {author} {\bibfnamefont {J.}~\bibnamefont {Rigby}}, \
  and\ \bibinfo {author} {\bibfnamefont {E.}~\bibnamefont {Daly}},\ }\href@noop
  {} {\bibfield  {journal} {\bibinfo  {journal} {Physical Review Letters}\
  }\textbf {\bibinfo {volume} {98}},\ \bibinfo {pages} {094101} (\bibinfo
  {year} {2007})}\BibitemShut {NoStop}%
\bibitem [{\citenamefont {Rorai}\ \emph {et~al.}(2015)\citenamefont {Rorai},
  \citenamefont {Mininni},\ and\ \citenamefont {Pouquet}}]{rorai2015stably}%
  \BibitemOpen
  \bibfield  {author} {\bibinfo {author} {\bibfnamefont {C.}~\bibnamefont
  {Rorai}}, \bibinfo {author} {\bibfnamefont {P.}~\bibnamefont {Mininni}}, \
  and\ \bibinfo {author} {\bibfnamefont {A.}~\bibnamefont {Pouquet}},\
  }\href@noop {} {\bibfield  {journal} {\bibinfo  {journal} {Physical Review
  E}\ }\textbf {\bibinfo {volume} {92}},\ \bibinfo {pages} {013003} (\bibinfo
  {year} {2015})}\BibitemShut {NoStop}%
\bibitem [{\citenamefont {Taylor}(1938)}]{taylor1938spectrum}%
  \BibitemOpen
  \bibfield  {author} {\bibinfo {author} {\bibfnamefont {G.~I.}\ \bibnamefont
  {Taylor}},\ }in\ \href@noop {} {\emph {\bibinfo {booktitle} {Proceedings of
  the Royal Society of London A: Mathematical, Physical and Engineering
  Sciences}}},\ Vol.\ \bibinfo {volume} {164}\ (\bibinfo {organization} {The
  Royal Society},\ \bibinfo {year} {1938})\ pp.\ \bibinfo {pages}
  {476--490}\BibitemShut {NoStop}%
\bibitem [{\citenamefont {Kraichnan}(1994)}]{kraichnan1994anomalous}%
  \BibitemOpen
  \bibfield  {author} {\bibinfo {author} {\bibfnamefont {R.~H.}\ \bibnamefont
  {Kraichnan}},\ }\href@noop {} {\bibfield  {journal} {\bibinfo  {journal}
  {Physical Review Letters}\ }\textbf {\bibinfo {volume} {72}},\ \bibinfo
  {pages} {1016} (\bibinfo {year} {1994})}\BibitemShut {NoStop}%
\bibitem [{\citenamefont {Prichard}\ and\ \citenamefont
  {Theiler}(1994)}]{prichard1994generating}%
  \BibitemOpen
  \bibfield  {author} {\bibinfo {author} {\bibfnamefont {D.}~\bibnamefont
  {Prichard}}\ and\ \bibinfo {author} {\bibfnamefont {J.}~\bibnamefont
  {Theiler}},\ }\href@noop {} {\bibfield  {journal} {\bibinfo  {journal}
  {Physical Review Letters}\ }\textbf {\bibinfo {volume} {73}},\ \bibinfo
  {pages} {951} (\bibinfo {year} {1994})}\BibitemShut {NoStop}%
\bibitem [{\citenamefont {Meneveau}(1991)}]{meneveau1991analysis}%
  \BibitemOpen
  \bibfield  {author} {\bibinfo {author} {\bibfnamefont {C.}~\bibnamefont
  {Meneveau}},\ }\href@noop {} {\bibfield  {journal} {\bibinfo  {journal}
  {Journal of Fluid Mechanics}\ }\textbf {\bibinfo {volume} {232}},\ \bibinfo
  {pages} {469} (\bibinfo {year} {1991})}\BibitemShut {NoStop}%
\bibitem [{\citenamefont {Li}\ and\ \citenamefont
  {Meneveau}(2005)}]{li2005origin}%
  \BibitemOpen
  \bibfield  {author} {\bibinfo {author} {\bibfnamefont {Y.}~\bibnamefont
  {Li}}\ and\ \bibinfo {author} {\bibfnamefont {C.}~\bibnamefont {Meneveau}},\
  }\href@noop {} {\bibfield  {journal} {\bibinfo  {journal} {Physical Review
  Letters}\ }\textbf {\bibinfo {volume} {95}},\ \bibinfo {pages} {164502}
  (\bibinfo {year} {2005})}\BibitemShut {NoStop}%
\bibitem [{\citenamefont {Mydlarski}\ and\ \citenamefont
  {Warhaft}(1998)}]{mydlarski1998passive}%
  \BibitemOpen
  \bibfield  {author} {\bibinfo {author} {\bibfnamefont {L.}~\bibnamefont
  {Mydlarski}}\ and\ \bibinfo {author} {\bibfnamefont {Z.}~\bibnamefont
  {Warhaft}},\ }\href@noop {} {\bibfield  {journal} {\bibinfo  {journal}
  {Journal of Fluid Mechanics}\ }\textbf {\bibinfo {volume} {358}},\ \bibinfo
  {pages} {135} (\bibinfo {year} {1998})}\BibitemShut {NoStop}%
\bibitem [{\citenamefont {Lepore}\ and\ \citenamefont
  {Mydlarski}(2009)}]{lepore2009effect}%
  \BibitemOpen
  \bibfield  {author} {\bibinfo {author} {\bibfnamefont {J.}~\bibnamefont
  {Lepore}}\ and\ \bibinfo {author} {\bibfnamefont {L.}~\bibnamefont
  {Mydlarski}},\ }\href@noop {} {\bibfield  {journal} {\bibinfo  {journal}
  {Physical review letters}\ }\textbf {\bibinfo {volume} {103}},\ \bibinfo
  {pages} {034501} (\bibinfo {year} {2009})}\BibitemShut {NoStop}%
\bibitem [{\citenamefont {Celani}\ \emph {et~al.}(2000)\citenamefont {Celani},
  \citenamefont {Lanotte}, \citenamefont {Mazzino},\ and\ \citenamefont
  {Vergassola}}]{celani2000universality}%
  \BibitemOpen
  \bibfield  {author} {\bibinfo {author} {\bibfnamefont {A.}~\bibnamefont
  {Celani}}, \bibinfo {author} {\bibfnamefont {A.}~\bibnamefont {Lanotte}},
  \bibinfo {author} {\bibfnamefont {A.}~\bibnamefont {Mazzino}}, \ and\
  \bibinfo {author} {\bibfnamefont {M.}~\bibnamefont {Vergassola}},\
  }\href@noop {} {\bibfield  {journal} {\bibinfo  {journal} {Physical Review
  Letters}\ }\textbf {\bibinfo {volume} {84}},\ \bibinfo {pages} {2385}
  (\bibinfo {year} {2000})}\BibitemShut {NoStop}%
\bibitem [{\citenamefont {Falkovich}(2009)}]{falkovich2009symmetries}%
  \BibitemOpen
  \bibfield  {author} {\bibinfo {author} {\bibfnamefont {G.}~\bibnamefont
  {Falkovich}},\ }\href {http://stacks.iop.org/1751-8121/42/i=12/a=123001}
  {\bibfield  {journal} {\bibinfo  {journal} {Journal of Physics A:
  Mathematical and Theoretical}\ }\textbf {\bibinfo {volume} {42}},\ \bibinfo
  {pages} {123001} (\bibinfo {year} {2009})}\BibitemShut {NoStop}%
\bibitem [{\citenamefont {Ozmidov}(1965)}]{ozmidov1965turbulent}%
  \BibitemOpen
  \bibfield  {author} {\bibinfo {author} {\bibfnamefont {R.}~\bibnamefont
  {Ozmidov}},\ }\href@noop {} {\bibfield  {journal} {\bibinfo  {journal}
  {Atmos. Oceanic Phys.}\ }\textbf {\bibinfo {volume} {1}},\ \bibinfo {pages}
  {861} (\bibinfo {year} {1965})}\BibitemShut {NoStop}%
\bibitem [{\citenamefont {Dougherty}(1961)}]{dougherty1961anisotropy}%
  \BibitemOpen
  \bibfield  {author} {\bibinfo {author} {\bibfnamefont {J.}~\bibnamefont
  {Dougherty}},\ }\href@noop {} {\bibfield  {journal} {\bibinfo  {journal}
  {Journal of Atmospheric and Terrestrial Physics}\ }\textbf {\bibinfo {volume}
  {21}},\ \bibinfo {pages} {210} (\bibinfo {year} {1961})}\BibitemShut
  {NoStop}%
\bibitem [{\citenamefont {Antonia}\ \emph {et~al.}(1984)\citenamefont
  {Antonia}, \citenamefont {Hopfinger}, \citenamefont {Gagne},\ and\
  \citenamefont {Anselmet}}]{antonia1984temperature}%
  \BibitemOpen
  \bibfield  {author} {\bibinfo {author} {\bibfnamefont {R.}~\bibnamefont
  {Antonia}}, \bibinfo {author} {\bibfnamefont {E.}~\bibnamefont {Hopfinger}},
  \bibinfo {author} {\bibfnamefont {Y.}~\bibnamefont {Gagne}}, \ and\ \bibinfo
  {author} {\bibfnamefont {F.}~\bibnamefont {Anselmet}},\ }\href@noop {}
  {\bibfield  {journal} {\bibinfo  {journal} {Physical Review A}\ }\textbf
  {\bibinfo {volume} {30}},\ \bibinfo {pages} {2704} (\bibinfo {year}
  {1984})}\BibitemShut {NoStop}%
\bibitem [{\citenamefont {Meneveau}\ \emph {et~al.}(1990)\citenamefont
  {Meneveau}, \citenamefont {Sreenivasan}, \citenamefont {Kailasnath},\ and\
  \citenamefont {Fan}}]{meneveau1990joint}%
  \BibitemOpen
  \bibfield  {author} {\bibinfo {author} {\bibfnamefont {C.}~\bibnamefont
  {Meneveau}}, \bibinfo {author} {\bibfnamefont {K.}~\bibnamefont
  {Sreenivasan}}, \bibinfo {author} {\bibfnamefont {P.}~\bibnamefont
  {Kailasnath}}, \ and\ \bibinfo {author} {\bibfnamefont {M.}~\bibnamefont
  {Fan}},\ }\href@noop {} {\bibfield  {journal} {\bibinfo  {journal} {Physical
  Review A}\ }\textbf {\bibinfo {volume} {41}},\ \bibinfo {pages} {894}
  (\bibinfo {year} {1990})}\BibitemShut {NoStop}%
\bibitem [{\citenamefont {Ruiz-Chavarria}\ \emph {et~al.}(1996)\citenamefont
  {Ruiz-Chavarria}, \citenamefont {Baudet},\ and\ \citenamefont
  {Ciliberto}}]{ruiz1996scaling}%
  \BibitemOpen
  \bibfield  {author} {\bibinfo {author} {\bibfnamefont {G.}~\bibnamefont
  {Ruiz-Chavarria}}, \bibinfo {author} {\bibfnamefont {C.}~\bibnamefont
  {Baudet}}, \ and\ \bibinfo {author} {\bibfnamefont {S.}~\bibnamefont
  {Ciliberto}},\ }\href@noop {} {\bibfield  {journal} {\bibinfo  {journal}
  {Physica D: Nonlinear Phenomena}\ }\textbf {\bibinfo {volume} {99}},\
  \bibinfo {pages} {369} (\bibinfo {year} {1996})}\BibitemShut {NoStop}%
\end{thebibliography}%

\end{document}